\documentclass[journal]{IEEEtran}

\makeatletter
\def\@IEEEtablestring{table}
\makeatother
\usepackage{cite}
\usepackage{amsmath,amssymb,amsfonts}
\usepackage{graphicx}
\usepackage{textcomp}
\usepackage{xcolor}
\usepackage{algpseudocode}
\usepackage{subfigure} 
\usepackage{algorithm2e}
\usepackage{amsthm}
\newtheorem{lemma}{Lemma}
\newtheorem{proposition}{Proposition}
\usepackage{url}        
\usepackage{breakurl}   
\usepackage{hyperref}   

\usepackage{booktabs}       
\usepackage{amsfonts}       
\usepackage{nicefrac}       
\usepackage{microtype}      

\title{Learning During Detection: Continual Learning for Neural OFDM Receivers via DMRS}

\author{
  Mohanad Obeed, \emph{Member, IEEE} and Ming Jian, \emph{Member, IEEE} \thanks{The authors are with the Advanced Wireless Technology Lab, Huawei Technologies Canada Co. Ltd,
ON K2K 3J1, Canada (e-mail: mohanad.obeed@huawei.com, ming.jian@gmail.com). Part of this work was presented at the IEEE International Conference on Communications (ICC), Glasgow, U.K., May 2026~\cite{obeed2026learning}.} \\
  
}
\begin{document}

\maketitle
\begin{abstract}
Deep neural networks (DNNs) have been increasingly explored for receiver
design because they can handle complex environments without relying on
explicit channel models. Nevertheless, because communication channels
change rapidly, their distributions can shift over time, often making
periodic retraining necessary. This paper proposes an online and
continual learning framework for orthogonal frequency-division
multiplexing (OFDM) neural receivers that directly detect the soft bits
of received signals. In two of its three pilot designs, the
framework consumes no data-carrying resource elements. Unlike conventional fine-tuning methods that rely on
dedicated training intervals or full resource grids, our approach
leverages existing demodulation reference signals (DMRS) to
simultaneously enable signal demodulation and model adaptation. We
introduce three pilot designs: fully randomized, hybrid, and additional
pilots that flexibly support joint demodulation and learning. To
accommodate these pilot designs, we develop two receiver architectures:
(i) a parallel design that separates inference and fine-tuning for
uninterrupted operation, and (ii) a forward-pass--reusing design that
reduces computational complexity.
We further show that the masked-pilot loss is an
unbiased estimator of the full-grid loss, derive the optimal masking
ratio and a condition under which the receiver tracks a drifting channel,
and quantify the computational cost of both architectures.
Simulation results show that the proposed method tracks both slow and
fast variations in delay spread and user velocity
without spectral overhead or service interruption, that
its error decreases monotonically across repeated traversals of the same
channel conditions, and that it remains stable where methods relying on
decoder-derived labels diverge at low signal-to-noise ratio or high code
rate.
\end{abstract}

\section{Introduction}
In wireless systems, the communication channel between transmitters and receivers is inherently uncontrollable as it continuously varies over time, frequency, and location. To ensure reliable data transmission, both ends of the link must be designed to cope with these fluctuations and reduce the uncertainty introduced by the channel. A common solution is to transmit known pilot signals at regular intervals, which the receiver uses to estimate the channel, equalize the received signal, and correctly demodulate the transmitted data. Typically, least squares (LS) methods are applied for channel estimation, followed by linear minimum mean square error (LMMSE) equalization. However, LS-LMMSE techniques cannot fully exploit temporal, spectral, or spatial correlations, and they lack the ability to identify or mitigate interference patterns.

Deep learning (DL) has recently emerged as a promising alternative for joint channel estimation and signal detection. DL-based receivers have demonstrated significant improvements in reducing bit-error rate (BER) in orthogonal frequency-division multiplexing (OFDM) systems \cite{honkala2021deeprx,korpi2021deeprx, ye2017power}. These performance gains come from neural networks’ ability to capture hidden correlations such as temporal dependencies and to recognize and mitigate noise and interference. Neural models can either replace individual receiver functions (e.g., channel estimation) or replace multiple functions such as estimation, equalization, and demapping into a unified framework.


Despite their strong performance, DL-based receivers often fail when the statistical properties of the operating channel deviate from those encountered during training \cite{9207745, farahani2021brief}. As demonstrated in \cite{luostari2025adapting}, models trained solely on simulated datasets can exhibit significant performance degradation when evaluated on real-world data. Because such models rely entirely on training data to learn patterns, they implicitly assume that the training distribution accurately represents the deployment environment. When factors such as fading characteristics, interference, or user mobility differ from those in the training data, the receiver’s performance deteriorates sharply.

This issue, known as distribution shift (or domain shift), occurs when the model fails to generalize beyond its training domain. In \cite{obeed2025hybrid}, it was shown that a neural receiver can even underperform a conventional receiver (e.g., LS-LMMSE) under distribution shifts. To mitigate this, the authors proposed a hybrid receiver architecture that integrates the neural and traditional receivers, using a learnable decider to select the most suitable receiver for a given input distribution. While this approach enhances robustness, the overall performance remains constrained by the traditional receiver’s limitations when the distribution shift is large. Therefore, achieving robust performance under varying channel conditions requires adaptation mechanisms such as online learning, transfer learning, domain adaptation, or retraining with more diverse datasets.

Several studies have investigated online fine-tuning to adapt neural receivers to distribution shifts. A common strategy alternates between training and inference: dedicated pilot symbols transmitted over multiple transmission time intervals (TTIs) are used for training, while subsequent TTIs carry data \cite{raviv2024adaptive}. While effective, this alternating scheme has three key drawbacks: it introduces high pilot overhead, interrupts user data transmission during training phases, and fails to track rapid channel variations that occur between updates.

Online learning approaches were proposed to fine-tune the neural networks at the receiver that are used for channel estimation \cite{luan2023channelformer, xu2024learning}, channel state information (CSI) feedback \cite{du2022robust, zhang2024continuous, cui2022unsupervised}, or signal detection \cite{khani2020adaptive, jiang2021ai, schibisch2018online, fischer2022adaptive}. For channel estimation, the authors of \cite{luan2023channelformer} introduced additional pilots with higher power compared to the demodulation pilots that can be used to fine-tune the neural network at the receiver. The increased power of the added pilots guarantees low channel estimation error for such pilots. The authors of \cite{wang2022learn} proposed to reduce the training data needed for new environments by leveraging knowledge from known environments. A few-shot learning approach was proposed, enabled by an attention-based method, to help the communication model generalize to new environments in a continual learning context.
The authors of \cite{xu2024learning} proposed an online learning approach where the channel coefficients are learned through the symbols detection loss.  They leveraged the repetitive pattern of modulation constellation and interference invariant properties to efficiently fine-tune the neural networks using reference signals. 

To reduce the training burden, compact models such as MMNet \cite{khani2020adaptive} were proposed, with only $\sim$40k parameters to enable efficient fine-tuning with fewer samples. Similarly, SwitchNet \cite{jiang2021ai} introduced fine-tuning of only a subset of parameters ($\alpha$’s), significantly reducing the number of training batches required. However, SwitchNet relies on offline-pretrained subnetworks, limiting adaptability to unseen channels. Although both MMNet and SwitchNet reduce complexity, they still require complete training samples, leading to non-negligible overhead and delays.

Other works \cite{schibisch2018online, fischer2022adaptive} proposed on-the-fly fine-tuning, where training labels are recovered using error-correcting codes rather than dedicated pilots. While this removes the need for extra pilots, it has two drawbacks: (i) including the decoder in the learning loop increases latency, and (ii) the decoder’s success depends on accurate log-likelihood ratios (LLRs) from the neural receiver, which are themselves vulnerable to degradation under distribution shift.

In this paper, we propose an online and continual learning approach for
neural OFDM receivers that adapts without interrupting inference and,
in two of its three pilot designs, without consuming data-carrying resource elements. The
key idea is to redesign demodulation pilots so they support both signal
demodulation and model fine-tuning. Three pilot designs are introduced:
(i) fully randomized learning-demodulation pilots, (ii) hybrid pilots
combining learning and conventional roles, and (iii) additional pilots
added alongside standard ones. To exploit these, two receiver
architectures are developed: one enabling parallel fine-tuning and
inference, and another that reuses the forward pass to reduce computation
but requires brief pauses during backpropagation.
In summary, this paper makes the following contributions:
\begin{itemize}
\item We introduce an online learning strategy for neural OFDM receivers,
where demodulation pilots are redesigned to serve simultaneously for
signal demodulation and neural fine-tuning.

\item We propose three flexible pilot designs: fully randomized
learning-demodulation pilots, hybrid pilots, and additional
pilots. These are not independent mechanisms but three
points on a compatibility--overhead trade-off, allowing deployment at
different stages of standards adoption.

\item We develop two receiver architectures that allow online fine-tuning
without interrupting inference: (i) fully parallel demodulation and
training, and (ii) a forward-pass--reusing design that reduces
computation.

\item We analyze the latency in the communication link that is introduced
by the DL model updates caused by the forward and backward passes.

\item We provide an analytical characterization of the
mechanism. We show that the masked-pilot loss is an unbiased estimator of
the loss obtained with complete label knowledge, with variance inversely
proportional to the number of masked pilots; we derive the optimal
masking ratio by balancing demodulation quality against gradient
variance; and we establish a condition relating the steady-state
adaptation error to the channel drift rate and the update interval.

\item We quantify the cost of the framework, separating
spectral, computational and standardization overhead, and report the
parameter count, FLOPs and amortized cost of both architectures as a
function of the update interval.

\item Through simulations, we demonstrate that the proposed approach
effectively tracks both slow and fast channel variations
in delay spread and user velocity without service
interruption or spectral overhead.
We further characterize the number of adaptation samples
required to restore performance after a channel change, show that the
error decreases monotonically across repeated traversals of the same
channel conditions, and compare against online learning based on
decoder-derived labels, which diverges at low signal-to-noise ratio or
high code rate where the proposed scheme remains stable.
\end{itemize}

A preliminary version of this work was presented
in~\cite{obeed2026learning}. The present paper extends it substantially;
in particular, the analytical characterization of
Section~\ref{sec:analysis}, the overhead and complexity analysis of
Section~\ref{sec:overhead}, and the studies of user velocity, recovery
sample complexity, decoder-derived labels, and uncertainty quantification
in Section~\ref{sec:results} are new.

\section{System model}
We consider an uplink communication system consisting of a single-antenna user transmitting to a base station equipped with $K$ receive antennas. OFDM is employed with $T$ OFDM symbols per frame and $F$ subcarriers. A random sequence of uniformly distributed information bits is generated at the transmitter. These bits are encoded using a low-density parity-check (LDPC) code and then converted into symbols. The symbols are then mapped into a two-dimensional resource grid (RG) of size $T \times F$ in the OFDM frame. A subset of resource elements is reserved for demodulation pilots, while the remaining elements carry data symbols. Traditionally, fixed pilots are placed at specific OFDM symbols (e.g., the 2nd and 11th symbols), spanning all subcarriers.  Before transmission, the inter-symbol interference is reduced by adding a cyclic prefix to the start of each OFDM symbol.

Let $Y \in \mathbb{C}^{T \times F \times K}$ denote the received resource grid at the base station. After removing the cyclic prefix and applying fast Fourier transform (FFT), each resource element can be expressed as
\begin{equation}
    y_{t,f,k} = h_{t,f,k}s_{t,f}+w_{t,f,k},
\end{equation}
where $s_{t,f}$ is the transmitted symbol located in the RG at the $t^{th}$ symbol and at the $f^{th}$ subcarrier, $h_{t,f,k}$ is the channel coefficient at antenna $k$, and $w_{t,f,k}$ is additive noise. 

Traditionally, channel estimation is performed using the received pilot symbols, followed by equalizing the frequency-domain signal using the estimated channel. The equalized symbols are then demapped into their corresponding constellation points. In contrast, our approach employs a single neural network that directly processes the input resource grids (RGs) to generate soft bits (LLRs), effectively replacing the conventional channel estimation, equalization, and demapping stages.
To form the input tensor, we concatenate the following along the feature dimension: Real and imaginary parts of $Y$, a pilot-symbol tensor $\mathbf{X}_{\mathrm{p}}$ $\in \mathbb{R}^{T \times F \times 2}$ (zeros at data positions, real/imaginary values at pilot positions), and the estimated noise variance. In the proposed approach, we need to mask some pilots, which means that its corresponding entry in $\mathbf{X}_{\mathrm{p}}$ is set to zero. The resulting input tensor has shape $\mathbb{R}^{T \times F \times C}$, where $C = 2(K +1)+1$, which allows the neural receiver to jointly exploit received samples, pilot information, and noise variance  \cite{honkala2021deeprx}. The network outputs LLRs, which are then passed to the channel decoder for final bit recovery.

\begin{figure*}[t!]
    \centering
        \includegraphics[scale=0.25]{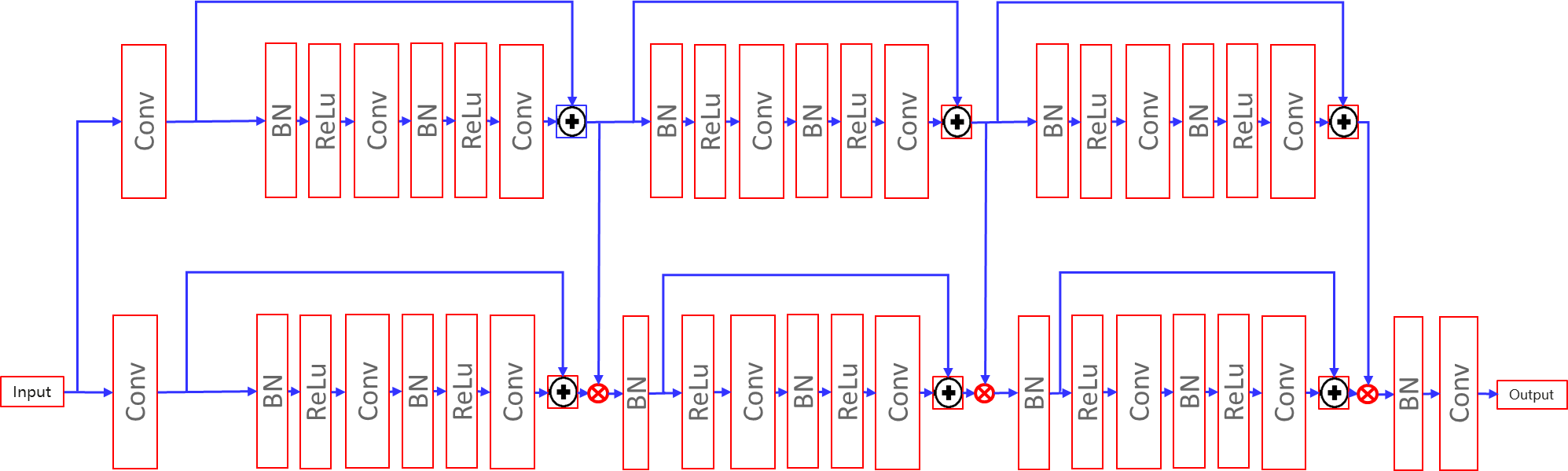}
    \caption{An example of CoNet architecture when depth = 3.}
    \label{fig:conet}
\end{figure*}

\subsection{Neural Network Architecture}
In this subsection, we present the neural network architecture employed in the proposed wireless receiver. A key requirement for the online learning framework is to select a neural network that not only achieves high detection accuracy but also adapts rapidly to distribution shifts during deployment. A faster adaptation reduces the overhead associated with model fine-tuning.
To this end, we evaluated two neural network architectures: ResNet (specifically DeepRx \cite{honkala2021deeprx}) and CoNet \cite{conet}. Our experiments showed that the proposed online learning approach is effective in both architectures; however, it performs better and converges faster when integrated with CoNet.

The CoNet architecture shares similarities with conventional CNNs and ResNets, but instead of a single processing path from input to output, CoNet employs multiple parallel sub-networks (subnets). These subnets interact through mathematical operations such as element-wise multiplication. The authors of \cite{conet} demonstrated that using two interacting subnets, a main subnet and a supporting subnet, significantly enhances detection performance compared to traditional ResNets. Moreover, they showed that a shallow CoNet can achieve performance comparable to a substantially deeper ResNet, enabling faster adaptation under dynamic channel conditions. Fig.~\ref{fig:conet} illustrates an example of the CoNet architecture adopted for both offline and online learning in this work.
CoNet is designed to preserve spatial features throughout its layers. The number of output channels in the final convolutional layer is set to $\log_2(Q)$, where $Q$ denotes the modulation order.


The model is initially trained offline using a simulated dataset designed for broad diversity and robust generalization. This dataset includes samples generated under various channel conditions, such as varying delay spreads, Doppler shifts, and different channel models. During deployment, if the receiver detects performance degradation, it triggers online fine-tuning by requesting the transmitter to send the proposed demodulation pilots. The receiver then incrementally fine-tunes the neural network until the performance meets the desired level.

\section{Demodulation Pilots Design}
\label{Sec:Pilots}
To enable online learning without dedicated training transmissions, we redesign the demodulation pilots at the transmitter so that they can serve a dual purpose: (i) aiding demodulation at the receiver, and (ii) providing training signals for fine-tuning the neural model. A key challenge is that pilots are normally used as inputs for demodulation, whereas in training they are treated as labels for loss computation. We address this by proposing flexible pilot structures and different receiver architectures that allow the same symbols to be exploited for both purposes.

To ensure effective fine-tuning, the learning-demodulation pilots need to  be randomized in both location and symbol values across the resource grid. By varying their placement and constellation points from mini-batch to mini-batch, these pilots closely mimic the behavior of real data symbols which leads to preventing the model from overfitting to fixed pilot patterns. The pilot locations and values follow a pseudo-random sequence whose generator and placement rule are fixed in advance and known to both ends, so both sides agree on them without any seed being exchanged over the air (see Section~\ref{sec:overhead}).

\begin{figure}[t!]
    \centering
    \subfigure[Fully randomized pilots.]{
        \includegraphics[width=0.55\linewidth]{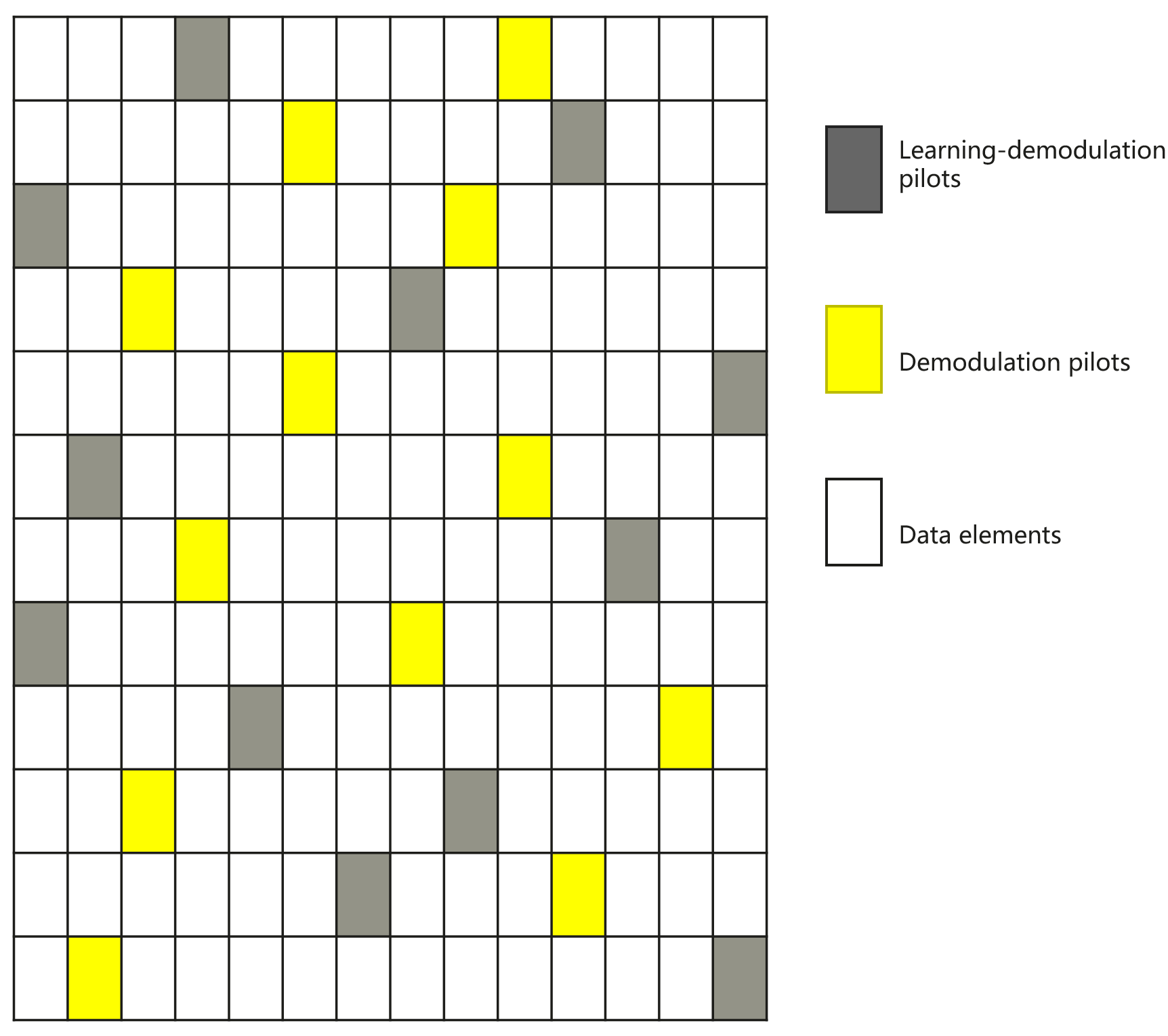}
        \label{fig:OFDM1}
    }
    \hfill
    \subfigure[Hybrid pilots.]{
        \includegraphics[width=0.55\linewidth]{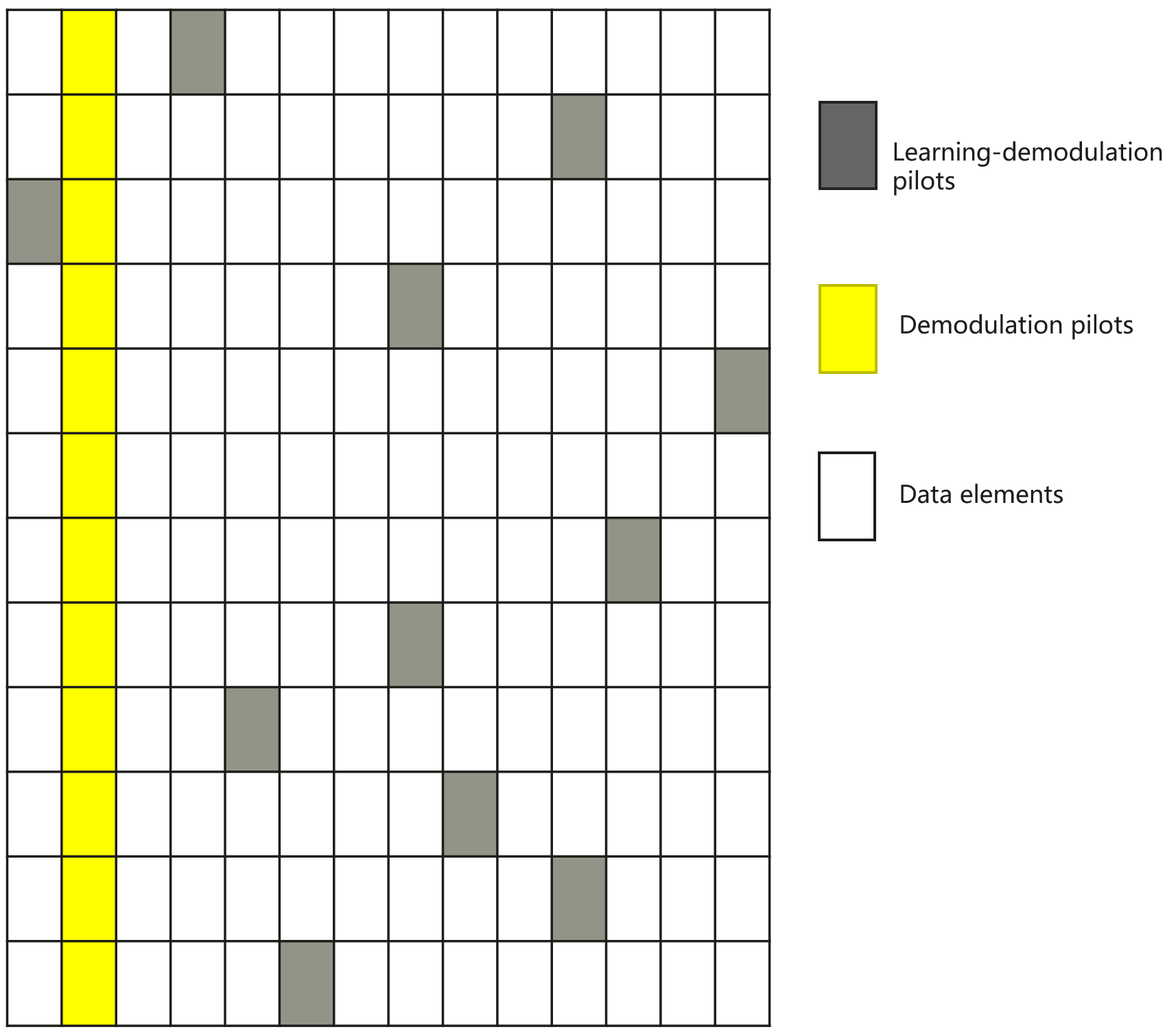}
        \label{fig:OFDM2}
        }
    \hfill
    \subfigure[Additional pilots for learning.]{
        \includegraphics[width=0.55\linewidth]{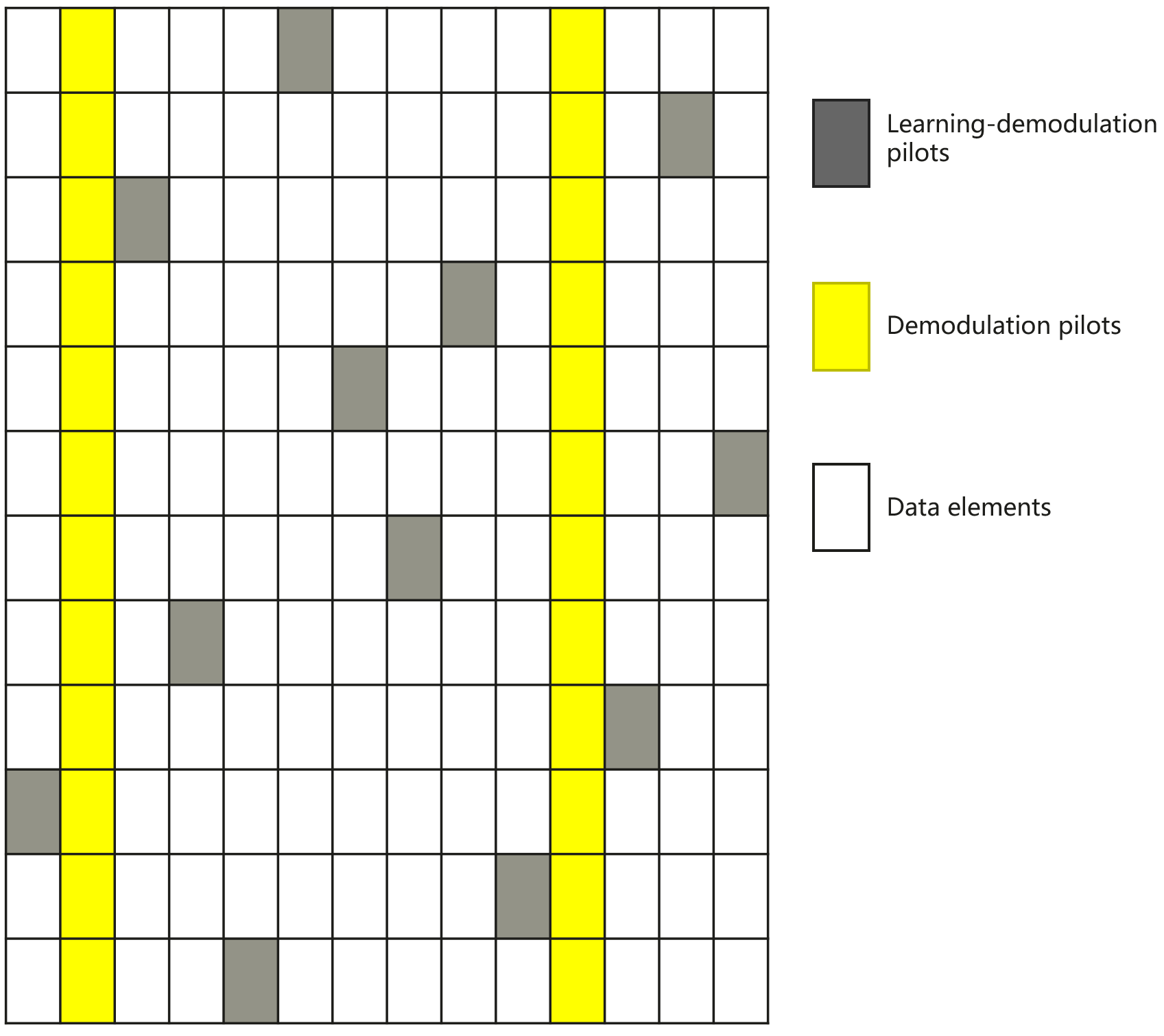}
        \label{fig:OFDM3}
        }
    \hfill
    \caption{Proposed OFDM frame structures.}
    \label{fig:Pilots}
\end{figure}
We consider three pilot designs, illustrated in Fig.~\ref{fig:Pilots}. The first design, shown in Fig.~\ref{fig:OFDM1}, uses fully randomized pilots. In this approach, pilot symbols are randomly drawn from the modulation constellation and scattered across arbitrary resource elements in the grid. Because they mimic the behavior of real data symbols, the network can use them both for demodulation and as labels for training. This design maximizes diversity and generalization, but it may complicate waveform design and synchronization.

The second design, shown in Fig.~\ref{fig:OFDM2}, introduces hybrid pilots. Here, a subset of pilots is randomized as in the first design and acts as \emph{learning-demodulation pilots}, while the remaining pilots follow conventional placement and modulation rules. This approach balances compatibility with existing pilot structures and flexibility for learning, but the number of pilots available for adaptation is reduced compared to the fully randomized case.

The third design, shown in Fig.~\ref{fig:OFDM3}, adds additional pilots. In this setup, extra symbols are transmitted alongside the standard demodulation pilots, and these additional pilots are used for both demodulation and training. This design is compatible with fixed pilot patterns but introduces a small amount of extra overhead, though still significantly less than conventional online fine-tuning.

Together, these designs provide a trade-off between backward compatibility and adaptation flexibility. Designs~(a) and~(b) consume no data-carrying resource
elements; design~(c) does, and the resulting trade-off is quantified in
Section~\ref{sec:overhead}.
In the following section, we show how different receiver architectures exploit these pilot designs to perform online learning during inference.
\section{Receiver design}
\label{sec:arch}
Traditionally, neural receivers operate in two separate modes: training (using known labels) and inference (detecting unknown data). This separation prevents the model from adapting continuously during operation. Our objective is to enable the receiver to learn from every inference sample while still delivering uninterrupted bit detection. 

We propose two receiver architectures that integrate online fine-tuning with real-time inference. Both architectures exploit the pilot designs introduced in Section \ref{Sec:Pilots}, but they differ in their balance between performance, latency, and computational complexity.
\begin{figure}[t!]
    \centering
    \subfigure[The proposed receiver: Architecture I.]{
        \includegraphics[width=0.8\linewidth]{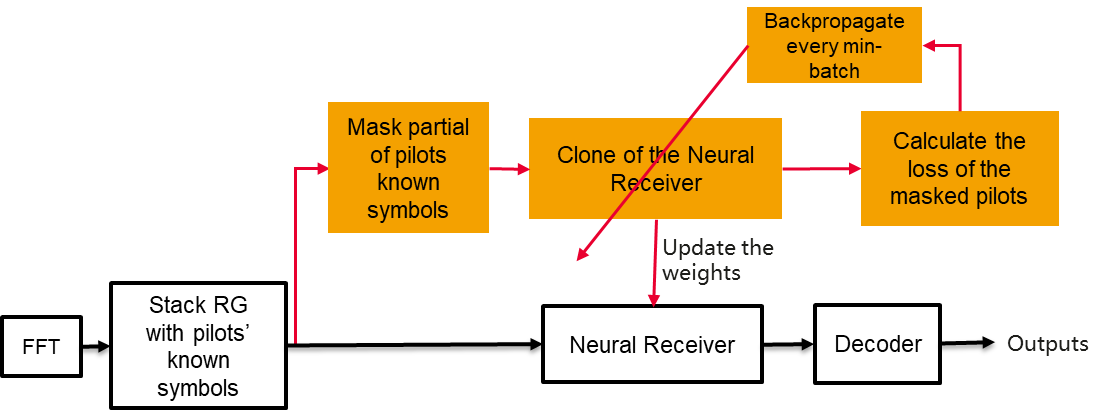}
        \label{fig:Rec1}
    }
    \hfill
    \subfigure[The proposed receiver: Architecture II.]{
        \includegraphics[width=0.8\linewidth]{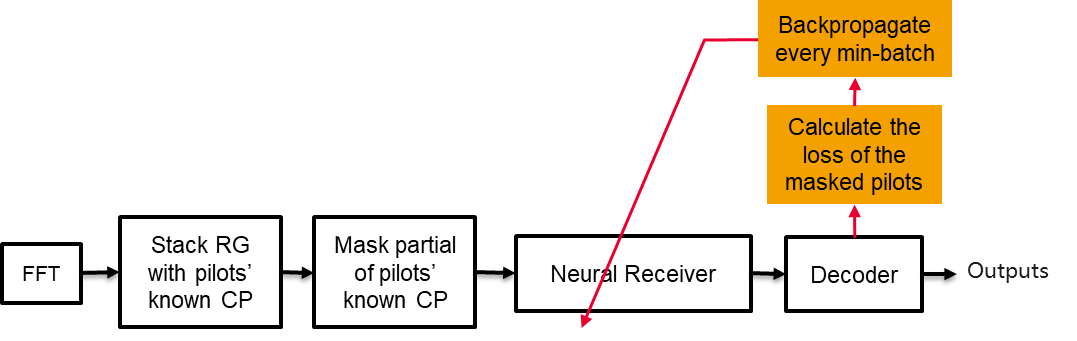}
        \label{fig:Rec2}
    }
    \hfill
    \caption{Proposed receivers that enable continual learning.}
    \label{fig:receivers}
\end{figure}
As shown in Fig.~\ref{fig:Rec1}, the receiver in Architecture I employs two parallel paths. The inference path receives the full RG with all pilots unmasked and produces LLRs for decoding. In parallel, a training path processes a masked version of the RG, where the symbols of a subset of the learning-demodulation pilots are set to zeros. The symbols of these masked pilots act as labels for computing the loss and performing backpropagation on a clone neural network. After each mini-batch, the clone network is updated, and its weights are copied to the inference model. This design avoids any interruption to inference and achieves strong performance since all pilots are available for demodulation, but it comes at the cost of higher computational complexity because two networks must be maintained.

In Fig.~\ref{fig:Rec2}, the receiver in Architecture II maintains only a single neural receiver. In this case, the input grid includes both unmasked pilots for demodulation and masked pilots for training. The forward pass is executed once to generate LLRs for both the data symbols and the masked pilots. After completing a mini-batch, the receiver computes the loss using the masked pilots and applies backpropagation directly to the same network. This approach reduces computational complexity since only one network is required, but it requires inference to pause during backpropagation and reduces the number of pilots available for demodulation.

These two architectures represent a trade-off: Architecture I prioritizes performance and robustness at the cost of computational resources, while Architecture II minimizes complexity but introduces limited performance degradation. Both designs, however, achieve the key goal of enabling online learning during inference using the existing pilot resources. The two architectures differ in computational cost
rather than in spectral cost; both are quantified in
Section~\ref{sec:overhead}.
\section{Delay analysis}
\label{Sec:Delay}
Here, we analyze the latency introduced by model updates resulting from forward and backward passes, focusing on the receiver with Architecture~I. Our goal is to show how frequently we can update the model to track the channel distribution shift and keep the performance stable.  
In communication systems, the end-to-end delay is dominated by the largest component (the bottleneck), such as the transmission or decoding delay. We divide the link latency into three parts:  
The \emph{pre-inference delay}, caused by operations prior to neural processing (e.g., transmission and time-domain processing),  
the \emph{inference delay} $I_d$, caused by the neural network forward pass used to produce LLRs, and  
the \emph{post-inference delay}, dominated by decoding operations and denoted $D_d$.  For simplicity, we denote the \emph{pre-inference delay} by $T_d$ since it is mostly dominated by the transmission delay and introduce $B_d$ as the delay from backpropagation during model fine-tuning. Our objective is to characterize the impact of $B_d$ on the receiver’s ability to adapt under channel distribution shifts.  

The computational cost of a linear layer is dominated by matrix multiplications. While the forward pass requires one multiplication, the backward pass involves two multiplications (for error propagation and gradient computation), leading to approximately double the cost of the forward pass \cite{goodfellow2016deep}. However, In practice, inference is performed sample-by-sample, whereas backpropagation is executed batch-by-batch, leveraging GPU parallelism. This implies that  $B_d$ of a mini-batch is equivalent to $2I_d$ of a sample. In some cases, when $I_d$ is the bottleneck in the communication link, the samples can be buffered and the inference process can be implemented in parallel for several samples at once. To generalize the relation between $B_d$ and $I_d$ in our system, let $N$ be the mini-batch size used for backpropagation, $Z$ the ratio between backpropagation delay and the inference delay for a single sample (empirically $Z \approx 2$ \cite{goodfellow2016deep}), and $M$ the number of samples processed in parallel during inference (typically $M = 1$, unless buffering is used). The general relation between $B_d$ and $I_d$ is then  
\[
B_d = Z \cdot \frac{I_d \cdot M}{N}.
\]  
 If $\max(T_d, D_d) < I_d$, the inference stage becomes the bottleneck. In this case, $I_d$ can be expressed as  
\[
I_d = V \cdot \max(T_d, D_d),
\]  
which allows $M = \lceil V \rceil$ samples to be buffered and processed in parallel. The maximum backpropagation delay occurs when $N=M$, yielding  
\[
B_d = Z \cdot I_d.
\]  
From this analysis, three operating cases arise: 
\begin{itemize}
    \item \textbf{Case I:} $\max(T_d, D_d) \geq I_d + B_d$. The model has sufficient time to update after every mini-batch. This is the best case since learning occurs continuously.  

 \item \textbf{Case II:} $\max(T_d, D_d) < I_d$ with $N \geq ZM$. The model updates every $\lceil 1 + \frac{ZM}{N} \rceil$ mini-batches. Since $(1 + \frac{ZM}{N}) \leq 2$, this leads to $\lceil 1 + \frac{ZM}{N} \rceil = 2$, which means that the  case here is updating every second mini-batch.  

\item \textbf{Case III:} $\max(T_d, D_d) < I_d$ with $N=M$. The model updates every third mini-batch: one forward pass plus two backward passes. 
\end{itemize}

Denote $U$ as the number of mini-batches required to update the model. In the simulations, we focus on Case II ($U=2$) and Case III ($U=3$), since Case I naturally provides superior performance.  

\section{Overhead and Complexity}
\label{sec:overhead}

The proposed framework has been described so far without quantifying
what it costs. We do so here, and we distinguish three forms of overhead
that are often conflated: spectral overhead, meaning resource elements
withdrawn from data transmission; computational overhead, meaning
additional arithmetic, latency and memory; and standardization impact,
meaning the extent to which the design departs from an existing
specification. The three pilot designs of Section~\ref{Sec:Pilots} and
the two architectures of Section~\ref{sec:arch} have different points
in each.

\subsection{Spectral Overhead and Standardization Impact}

Designs~(a) and~(b) of Fig.~\ref{fig:Pilots} reuse pilots that would already be transmitted, modifying only their values and, for Design~(a), their positions. Therefore, neither design allocates additional resource elements to pilots, and both incur zero spectral overhead. In contrast, Design~(c) introduces $P_{\mathrm{a}}$ additional pilot resource elements, resulting in a spectral overhead of $P_{\mathrm{a}}/N_{\mathrm{RE}}$, where $N_{\mathrm{RE}}$ is the number of resource elements per grid. For the configuration considered in this work, this corresponds to $7.1\%$ of the available resource elements. Despite this additional overhead, Design~(c) is included because it preserves the existing DMRS structure and can therefore be deployed without requiring changes to the DMRS-related standards. Importantly, the additional pilots are not used solely as labels for online learning. Their positions are also randomized in the same manner as the DMRS pilots in Design~(a), and the receiver jointly exploits both the original and additional pilots for demodulation. Consequently, the additional pilots also contribute to channel estimation, partially offsetting their spectral cost through improved demodulation performance.

Regarding the coordination of pilot values, the transmitter and receiver
must agree on the sequence from which the randomized pilots are drawn.
We regard this as a standardization matter rather than a per-transmission
overhead: the generating sequence and the placement rule can be fixed in
the specification and are therefore known to both ends in advance. No
sequence is exchanged over the air, no additional control signaling is
introduced, and no per-slot negotiation takes place. The only operational
requirement is that the two ends remain synchronized on the sequence
index, which is of the same nature as the synchronization any
pilot-based receiver already maintains. We further note that the masking
operation itself requires no coordination whatsoever: which pilots are
withheld from the network input is decided locally at the receiver, and
the transmitter is unaffected by that choice.
Table~\ref{tab:overhead} summarizes the three designs.

\begin{table}[!t]
\renewcommand{\arraystretch}{1.2}
\caption{Overhead of the Three Pilot Designs.}
\label{tab:overhead}
\centering
\footnotesize
\begin{tabular}{@{}lccc@{}}
\toprule
 & \textbf{(a) Fully} & \textbf{(b) Hybrid} & \textbf{(c) Additional} \\
 & \textbf{randomized} & & \textbf{pilots} \\
\midrule
Data REs consumed          & none & none & $7.1\%$ \\
Extra pilots aid demod.    & ---  & ---  & yes \\
\midrule
DMRS values changed        & yes & yes & no \\
DMRS positions changed     & yes & partly & no \\
Backward compatible        & no & partly & yes \\
\midrule
Over-the-air signaling    & none & none & none \\
Pre-agreed sequence needed & yes & yes & yes \\
Masking decided at Rx only & yes & yes & yes \\
\bottomrule
\end{tabular}
\end{table}

\subsection{Computational Overhead}

The receiver has $606\,854$ parameters, occupying $2.43$~MB in single
precision, and requires $1.078$~GFLOP per resource grid for one forward
pass, counting one multiply--accumulate as two operations. We take the
backward pass as twice the forward pass, which is the usual convention.

The peak cost is higher than a doubling. When an update is performed
under Architecture~I, the detection copy performs its usual forward pass
while the learning copy performs an additional forward pass and a
backward pass, so an update slot costs four times a fixed neural
receiver. Architecture~II reuses the detection forward pass as the
forward pass of the learning step, so an update slot costs three times
rather than four; it also does not clone the network, and its parameter
count and weight memory are therefore identical to those of a fixed
receiver. The doubling of weight memory applies to Architecture~I alone.

This peak cost is not paid on every slot. The additional computation is
incurred only when an update takes place, and by the analysis of
Section~\ref{Sec:Delay} an update occurs once every $U$ mini-batches.
Averaged over $U$, the cost relative to a fixed neural receiver is
$1+3/U$ for Architecture~I and $1+2/U$ for Architecture~II. The update
interval therefore governs the computational cost as well as the
adaptation delay, and the two may be traded against one another using a
single parameter. Table~\ref{tab:complexity} reports the resulting
figures together with the measured latency.

Memory follows the same pattern. Architecture~I stores a second copy of
the network, increasing the weight memory from $2.43$~MB to $4.85$~MB,
whereas Architecture~II leaves it unchanged. On update slots both
architectures must additionally retain the layer activations required by
the backward pass. This cost is common to the two architectures, is
substantially larger than the weight storage itself, and scales linearly
with the batch size; the dominant memory cost of online adaptation is
therefore the backward pass rather than the duplication of the network.
It may be reduced by adopting a smaller batch, at the expense of a
noisier gradient and hence slower adaptation by
Lemma~\ref{lem:unbiased}. Together with $U$, this provides two
independent parameters by which computation and memory may be traded
against adaptation speed.

\begin{table}[!t]
\renewcommand{\arraystretch}{1.2}
\caption{Complexity of the Two Architectures Relative to a Fixed Neural Receiver}
\label{tab:complexity}
\centering
\footnotesize
\begin{tabular}{@{}lccc@{}}
\toprule
& \textbf{Fixed} & \textbf{Arch.\ I} & \textbf{Arch.\ II} \\
& \textbf{neural Rx} & \textbf{(parallel)} & \textbf{(reused)} \\
\midrule
\multicolumn{4}{@{}l}{\textit{Static cost}}\\
Clones the network              & ---     & yes      & no \\
Parameters ($\times 10^3$)      & $606.9$ & $1213.7$ & $606.9$ \\
Weight memory (MB)              & $2.43$  & $4.85$   & $2.43$ \\
\midrule
\multicolumn{4}{@{}l}{\textit{Cost of one slot} (GFLOP)}\\
Detection only                  & $1.08$ & $1.08$ & $1.08$ \\
Detection $+$ update            & ---    & $4.31$ & $3.23$ \\
\;\;\emph{relative to fixed Rx} & $1.00\times$ & $4.00\times$ & $3.00\times$ \\
\midrule
\multicolumn{4}{@{}l}{\textit{Amortized} (\,$\times$ fixed Rx\,)}\\
$U=1$                           & $1.00$ & $4.00$ & $3.00$ \\
$U=2$                           & $1.00$ & $2.50$ & $2.00$ \\
$U=3$                           & $1.00$ & $2.00$ & $1.67$ \\
\midrule
\multicolumn{4}{@{}l}{\textit{Measured latency} (ms per resource grid)}\\
Detection slot                  & $0.246$ & $0.246$ & $0.246$ \\
Update slot                     & ---     & $0.618$ & $0.371$ \\
Amortized, $U=2$                & $0.246$ & $0.432$ & $0.309$ \\
\bottomrule
\end{tabular}

\vspace{2pt}
\begin{minipage}{\columnwidth}
\footnotesize
FLOPs are per resource grid, counting one
multiply--accumulate as two operations and the backward pass as twice
the forward pass. Amortized rows assume one
update every $U$ mini-batches.
\end{minipage}
\end{table}

\section{Theoretical Analysis}
\label{sec:analysis}

This section explains why the proposed mechanism works. We first show
that the masked-pilot loss is an unbiased estimator of the objective the
receiver is ultimately evaluated on (Lemma~\ref{lem:unbiased}). We then
use its variance to derive the optimal masking ratio
(Proposition~\ref{prop:rho}) and to state a condition under which the
receiver tracks a drifting channel (Proposition~\ref{prop:tracking}).

\subsection{The Masked-Pilot Loss is Unbiased}
\label{sec:unbiased}

Let $\mathcal{G}$ denote the set of resource elements in one grid,
$|\mathcal{G}| = N_{\mathrm{RE}}$, and let $\ell_u(\theta)$ be the
per-element binary cross-entropy at element $u$ for receiver parameters
$\theta$. The quantity of interest is the full-grid loss
\begin{equation}
\mathcal{L}(\theta) \;=\; \frac{1}{N_{\mathrm{RE}}}
\sum_{u \in \mathcal{G}} \ell_u(\theta),
\label{eq:fullloss}
\end{equation}
which cannot be evaluated online, since the transmitted data symbols are
unknown at the receiver. The receiver instead evaluates
\begin{equation}
\widehat{\mathcal{L}}(\theta) \;=\; \frac{1}{|\mathcal{M}|}
\sum_{u \in \mathcal{M}} \ell_u(\theta),
\label{eq:maskedloss}
\end{equation}
where $\mathcal{M} \subset \mathcal{G}$ is the set of masked pilots,
$|\mathcal{M}| = \rho P$, with $P$ the number of learning-demodulation
pilots per grid and $\rho \in (0,1)$ the masking ratio.

\begin{lemma}[Unbiasedness and variance]
\label{lem:unbiased}
If the masked pilots are drawn i.i.d.\ from the same constellation as the
data symbols and are placed uniformly at random over the resource
elements not occupied by demodulation pilots, then
\begin{equation}
\mathbb{E}\!\left[\widehat{\mathcal{L}}(\theta)\right]
= \mathcal{L}(\theta),
\qquad
\mathrm{Var}\!\left[\widehat{\mathcal{L}}(\theta)\right]
= \frac{\sigma_\ell^2}{\rho P},
\end{equation}
where $\sigma_\ell^2$ is the variance of $\ell_u(\theta)$ across
resource elements. Consequently, for a mini-batch of $N$ grids, the
stochastic gradient
$\nabla_\theta \widehat{\mathcal{L}}$ is an unbiased estimator of
$\nabla_\theta \mathcal{L}$ with variance
$\sigma_g^2 = \sigma_0^2 / (N\rho P)$, where $\sigma_0^2$ is the variance of the
per-element gradient $\nabla_\theta \ell_u(\theta)$ across resource elements.
\end{lemma}

\begin{IEEEproof}
Uniform placement makes each $u \in \mathcal{M}$ a uniform draw from
$\mathcal{G}$, so $\mathbb{E}[\ell_u] = \mathcal{L}(\theta)$ for every
$u \in \mathcal{M}$, and linearity of expectation gives the first
result. Independence across the $\rho P$ draws gives the second. The
gradient statement follows by applying the same argument to
$\nabla_\theta \ell_u$ and averaging over $N$ independent grids.
\end{IEEEproof}

Lemma~\ref{lem:unbiased} is the reason that no specialized update rule
is required. The receiver performs ordinary stochastic gradient descent
on an unbiased estimate of the objective it is evaluated on; the only
consequence of observing a subset of the grid is a variance penalty that
decreases as $1/(\rho P)$.

We emphasize a property that distinguishes this from self-training and
test-time adaptation schemes. The labels used in \eqref{eq:maskedloss}
are symbols the transmitter has sent and the receiver already knows.
Their accuracy is therefore independent of $\theta$. In schemes that
construct labels from the receiver's own output - by re-encoding
decoded bits, or by rewarding confident predictions - label accuracy
is a function of $\theta$, so a degradation in $\theta$ degrades the
labels, which degrades $\theta$ further. This feedback is absent here by
construction, and is the reason plain fine-tuning remains stable.

\subsection{Optimal Masking Ratio}
\label{sec:rho}

Increasing $\rho$ has two opposing effects. It reduces the number of
demodulation pilots to $(1-\rho)P$, degrading channel estimation, and it
increases the number of labels to $\rho P$, reducing gradient variance.
We model the resulting excess error as the sum of the two contributions.

Channel-estimation error variance scales inversely with pilot density,
$\sigma_h^2 \propto 1/\big((1-\rho)P\big)$. For a loss that is locally
$\mu$-strongly convex about the tracking point, the steady-state excess
error of online SGD with step size $\eta$ is
$\eta \sigma_g^2 / (2\mu)$, which by Lemma~\ref{lem:unbiased} scales as
$1/(\rho P)$. Writing the two proportionality constants as $a$ and $b$,
\begin{equation}
\mathcal{E}(\rho) \;\approx\; \frac{a}{1-\rho} \;+\; \frac{b}{\rho}.
\label{eq:excess}
\end{equation}

\begin{proposition}[Optimal masking ratio]
\label{prop:rho}
The minimizer of \eqref{eq:excess} over $\rho \in (0,1)$ is
\begin{equation}
\rho^\star \;=\; \frac{\sqrt{b}}{\sqrt{a}+\sqrt{b}}
\;=\; \frac{1}{1+\sqrt{a/b}},
\label{eq:rhostar}
\end{equation}
and $\rho^\star = 1/2$ when $a = b$. Moreover
$\mathcal{E}''(\rho^\star) > 0$ with
$\mathcal{E}(\rho)/\mathcal{E}(\rho^\star) = 1 + O\big((\rho -
\rho^\star)^2\big)$, so the penalty for deviating from $\rho^\star$ is
second order.
\end{proposition}

\begin{IEEEproof}
Differentiating \eqref{eq:excess} gives
$\mathcal{E}'(\rho) = a/(1-\rho)^2 - b/\rho^2$. Setting this to zero
yields $\sqrt{a}\,\rho = \sqrt{b}\,(1-\rho)$ and hence
\eqref{eq:rhostar}. The second derivative
$\mathcal{E}''(\rho) = 2a/(1-\rho)^3 + 2b/\rho^3$ is positive on
$(0,1)$, so the stationary point is the unique minimum and the local
expansion is quadratic.
\end{IEEEproof}

Proposition~\ref{prop:rho} makes two predictions that we test in
Section~\ref{sec:results}: that the error as a function of $\rho$ is
U-shaped with an interior minimum near $\rho = 1/2$, and that
performance is insensitive to $\rho$ over a broad neighborhood of the
optimum. Fig.~\ref{fig:rho_sweep} confirms both. Fitting
\eqref{eq:excess} to the measured curve gives $\rho^\star = 0.47$,
consistent with the observed minimum at $\rho = 0.5$. We present
\eqref{eq:excess} as a first-order description of the trade-off rather
than a quantitative predictor: it captures the existence, location and
flatness of the optimum, but underestimates the curvature at the
extremes of the range, where effects not represented in the model become
significant.

Equation~\eqref{eq:rhostar} also indicates how $\rho^\star$ should move
with operating conditions. A more rapidly varying channel increases the
weight $b$ attached to adaptation quality and therefore raises
$\rho^\star$, while a receiver operating with a large pilot budget $P$
can afford a smaller $\rho$ for the same absolute number of labels.

\subsection{Tracking Condition}
\label{sec:tracking}
We now explain how the receiver adapts to a changing channel. Let $\theta^\star_t$ denote the minimizer of
$\mathcal{L}$ under the channel statistics at mini-batch $t$, and define
the drift
$\delta_t = \lVert \theta^\star_{t+1} - \theta^\star_t \rVert$. Suppose
$\mathcal{L}$ has $L$-Lipschitz gradients and the receiver updates once
every $U$ mini-batches with step size $\eta$.

\begin{proposition}[Steady-state tracking error]
\label{prop:tracking}
For $\eta \le 1/L$, the averaged squared gradient norm over $J$ updates
satisfies
{\small
\begin{equation}
\frac{1}{J}\sum_{t=1}^{J}
\mathbb{E}\big\lVert \nabla\mathcal{L}(\theta_t) \big\rVert^2
\le
\underbrace{\frac{2\big(\mathcal{L}(\theta_0)-\mathcal{L}^\star\big)}
{\eta J}}_{\text{transient}}
+
\underbrace{\eta L \sigma_g^2}_{\text{gradient noise}}
+
\underbrace{\frac{L^2 U^2 \bar{\delta}^2}{\eta}}_{\text{tracking lag}},
\label{eq:tracking}
\end{equation}
}
where $\bar{\delta}$ is the mean per-mini-batch drift of the optimum and
$\sigma_g^2 = \sigma_0^2/(N\rho P)$ by Lemma~\ref{lem:unbiased}.
\end{proposition}

\emph{Origin of the three terms.} The first two terms are the standard result
for stochastic gradient descent on a smooth non-convex objective with a
fixed optimum \cite{ghadimi2013}: the transient term measures how far
the receiver starts from a stationary point and vanishes as $J$ grows,
while the gradient-noise term is the floor imposed by estimating the
gradient from finitely many labels. Lemma~\ref{lem:unbiased} makes this
floor explicit for the proposed scheme, since $\sigma_g^2$ is inversely
proportional to the number of masked pilots used per update.

The third term is what the time-varying channel adds. Because the
receiver updates once every $U$ mini-batches, the optimum may move by up
to $U\bar{\delta}$ between consecutive updates, and the receiver is
always correcting towards where the channel was rather than where it now
is. The resulting penalty is quadratic in $U\bar{\delta}$ and inversely
proportional to the step size. The treatment of drifting
optima in modern time-varying optimization is surveyed in
\cite{simonetto2020}. 

\emph{Interpretation.} The second and third terms move in opposite
directions with the step size: a larger $\eta$ suppresses the lag but
raises the noise floor. The two are balanced when
\begin{equation}
\eta L \sigma_g^2 \;=\; \frac{L^2 U^2 \bar{\delta}^2}{\eta}
\qquad\Longrightarrow\qquad
\eta^\star \;=\; U\bar{\delta}\sqrt{\frac{L}{\sigma_g^2}},
\label{eq:etastar}
\end{equation}
which shows that a faster-drifting channel or a longer update interval
calls for a larger learning rate, while a noisier gradient - that is,
a smaller $\rho P$ - calls for a smaller one. Substituting
$\eta^\star$ back into \eqref{eq:tracking} gives a steady-state error of
order $U\bar{\delta}\,L^{3/2}\sigma_g$. This balance is attainable only if
$\eta^\star$ respects the stability limit $\eta \le 1/L$, which requires
\begin{equation}
U\bar{\delta} \;\lesssim\; \frac{1}{L^{3/2}}\sqrt{\frac{\sigma_0^2}{N\rho P}} \, ,
\label{eq:condition}
\end{equation}
beyond which the tracking-lag term dominates at every admissible step
size and the error grows quadratically in $U\bar{\delta}$. Condition
\eqref{eq:condition} is the quantitative form of the requirement anticipated
in Section~\ref{Sec:Delay}: shortening the update interval $U$ enlarges the
range of drift rates the receiver can follow. The
update interval $U$ therefore governs three quantities at once: the
adaptation delay analyzed in Section~\ref{Sec:Delay}, the computational
cost tabulated in Table~\ref{tab:complexity}, and the tracking lag in
\eqref{eq:tracking}. Cases~I--III correspond to three points on this
trade-off.

\subsection{Why Plain Fine-Tuning is Stable}
\label{sec:whystable}

Online fine-tuning during deployment is often unstable, and it is worth
stating why it is well behaved here. The reason is the origin of the
labels.

Existing online methods must construct their training targets from the
receiver's own output, either by re-encoding decoded bits
\cite{schibisch2018online, 8999801,  9242305} or by rewarding
confident predictions \cite{wang2021tent}. In such schemes the accuracy
of the labels is a function of the accuracy of the model. When the
channel changes and the receiver becomes mismatched, the labels degrade
at the same moment, and an update computed from degraded labels degrades
the receiver further. The resulting feedback is self-reinforcing, and it
is strongest precisely when adaptation is most needed. Our comparison in
Section~\ref{sec:ecc} shows this directly: below
$E_b/N_0 = 8$~dB the recovered labels of the decoder-based baseline
become unreliable and its bit error rate rises above that of a receiver
performing no adaptation at all.

In the proposed design the labels are pilot symbols that the transmitter
has sent and the receiver already knows. Their accuracy does not depend
on $\theta$, so the path from model error to label error does not exist.
Whatever the state of the receiver, each update is computed from correct
targets, and the receiver is always pulled towards the correct solution
rather than away from it.

Two further properties keep the updates small and well conditioned.
First, by Lemma~\ref{lem:unbiased} the masked-pilot loss is an unbiased
estimate of the full-grid loss, so the receiver descends the objective it
is actually evaluated on rather than a surrogate; ordinary stochastic
gradient descent is therefore sufficient and no specialized update rule
is required. Second, a change in delay spread or user velocity alters the
statistics of the receiver's input but not the mapping it must implement.
Adaptation is thus a refinement of an already near-optimal solution
rather than a re-training, which is also why little of what has been
learned is lost, as the revisit analysis of
Fig.~\ref{fig:forget} confirms.

Finally, because updates occur every $U$ mini-batches with $U$ small,
each update begins close to the solution it must reach. The framework
tracks a slowly moving optimum rather than repeatedly re-solving the
problem, and Proposition~\ref{prop:tracking} states the condition under
which this tracking succeeds.

\section{Simulation results}
\label{sec:results}
\begin{table}[!t]
\renewcommand{\arraystretch}{1.15}
\setlength{\tabcolsep}{3pt}
\caption{Experimental Settings and Hyperparameters}
\label{tab:settings}
\centering
\scriptsize
\begin{tabular}{@{}llll@{}}
\toprule
\textbf{Parameter} & \textbf{Value} & \textbf{Parameter} & \textbf{Value} \\
\midrule
\multicolumn{2}{@{}l}{\textit{OFDM system}}
  & \multicolumn{2}{l}{\textit{Neural receiver (CoNet)}} \\
Subcarriers        & $64$              & Conv.\ layers      & $19$ \\
OFDM symbols       & $14$              & Residual blocks    & $4$ per branch \\
Subcarrier spacing & $30$~kHz          & Filters per layer  & $64$ \\
Cyclic prefix      & $6$               & Kernel size        & $3\times3$ \\
Slot duration      & $0.5$~ms          & Normalization      & BatchNorm \\
REs per grid       & $896$             & Activation         & ReLU \\
Modulation         & $64$-QAM          & Input channels     & $7$ \\
Channel code       & 5G NR LDPC        & Output channels    & $6$ \\
Code rates         & $0.5$, $0.7$      & Parameters         & $606\,854$ \\
Tx / Rx antennas   & $1$ / $2$         & Weight memory      & $2.43$~MB \\
\midrule
\multicolumn{2}{@{}l}{\textit{Pilots}}
  & \multicolumn{2}{l}{\textit{Offline pre-training}} \\
Total pilot REs    & $128$             & Optimizer          & Adam \\
Masking ratio $\rho$ & $0.5$           & Learning rate      & $10^{-3}$ \\
$\rho$ examined    & $0.25$--$0.75$    & Batch size         & $16$ \\
Pilot values       & i.i.d.\ $64$-QAM  & $E_b/N_0$ range    & $7$--$10$~dB \\
Placement          & re-randomized     & Loss               & masked-pilot BCE \\
\midrule
\multicolumn{2}{@{}l}{\textit{Channel}}
  & \multicolumn{2}{l}{\textit{Online adaptation}} \\
Model              & 3GPP TDL-C        & Optimizer          & Adam \\
DS, static exps.   & $40$--$50$~ns     & Learning rate      & $10^{-4}$ \\
DS, dynamic range  & $50$--$500$~ns    & LRs examined       & $10^{-5}$--$10^{-3}$ \\
DS step            & $20$~ns / $30$ mb & Grad.\ clipping    & norm $1.0$ \\
Cycle period       & $1380$ mb         & Batch size         & $16$ \\
User speed         & $0$--$30$~m/s     & Update interval $U$ & $1$ / $2$ / $3$ \\
Normalization      & per-realization   & Run length         & $10\,000$ mb \\
\midrule
\multicolumn{2}{@{}l}{\textit{Evaluation}}
  & \multicolumn{2}{l}{\textit{Software}} \\
Independent runs   & $3$ (sample-complexity study)              & Framework          & Sionna  \\
Smoothing window   & $51$ mb           & Backend            & TensorFlow \\

\bottomrule
\end{tabular}

\vspace{2pt}
\begin{minipage}{\columnwidth}
\scriptsize
``mb'' denotes mini-batches of $16$ resource grids. Seeds
are fixed and shared across compared methods, so all receivers observe
identical channel and noise realizations at every mini-batch.
\end{minipage}
\end{table}

\begin{figure}[t!]
    \centering
    \subfigure[The receiver is tested on unseen samples drawn from the same distribution.]{
        \includegraphics[width=0.7\linewidth]{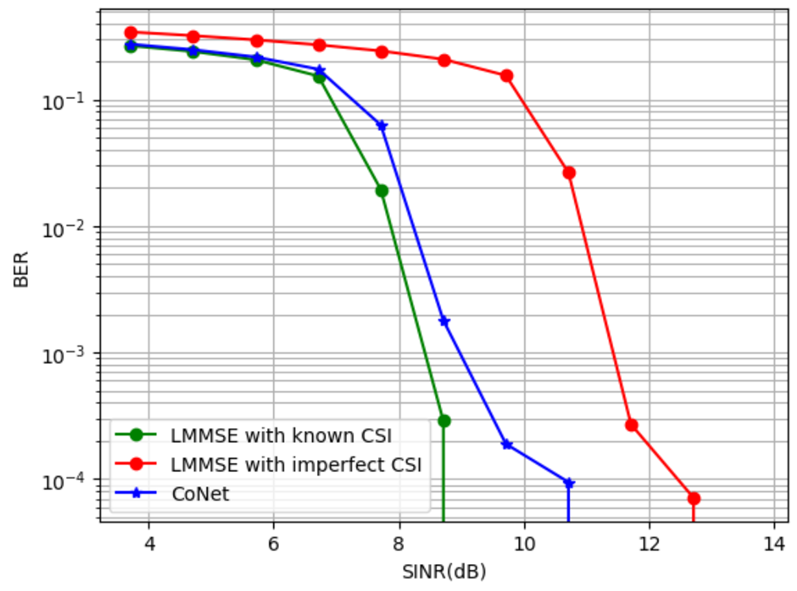}
        \label{fig:1}
    }
    \hfill
    \subfigure[The receiver is tested on unseen samples drawn from the drifted channel distribution.]{
        \includegraphics[width=0.7\linewidth]{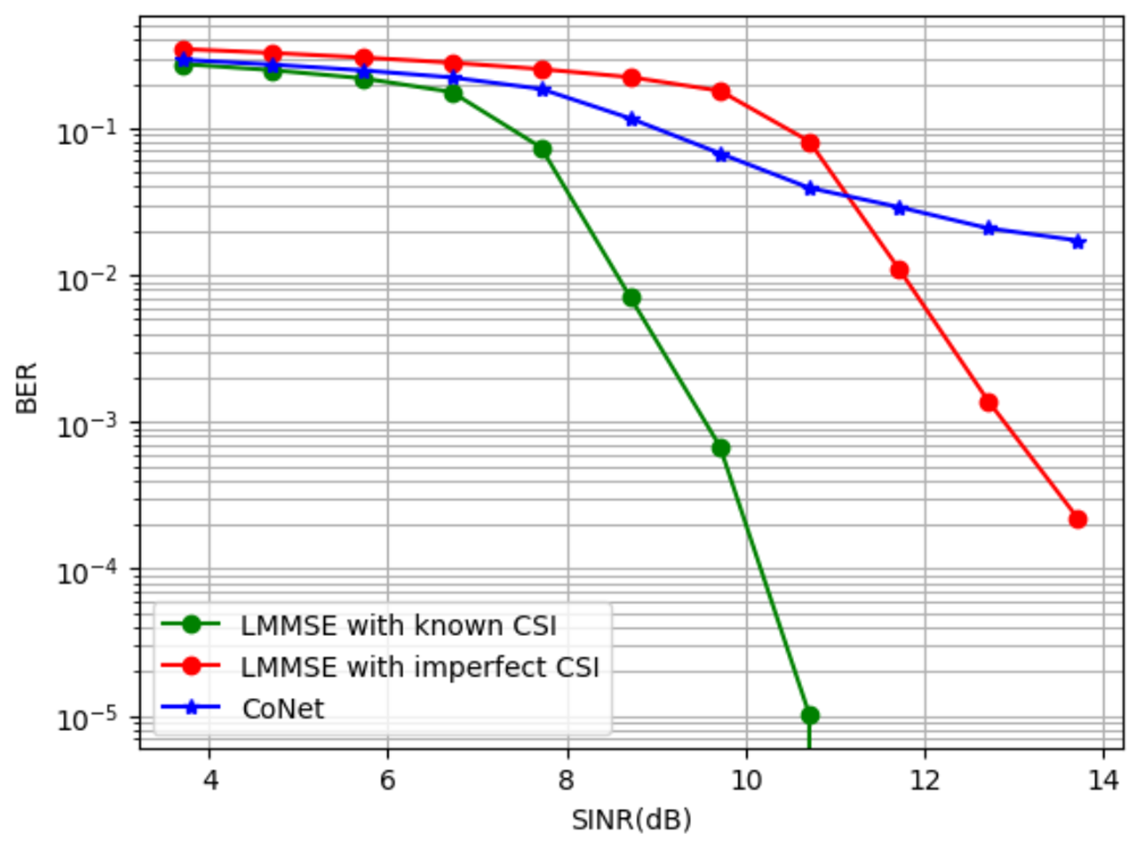}
        \label{fig:2}
    }
    \hfill
    \caption{BER versus SNR to show the effect of distribution shift.}
    \label{fig:combined22}
\end{figure}

\begin{figure}[t!]
    \centering
    \subfigure[The receiver is updated by 24K samples from the shifted distribution.]{
        \includegraphics[width=0.7\linewidth]{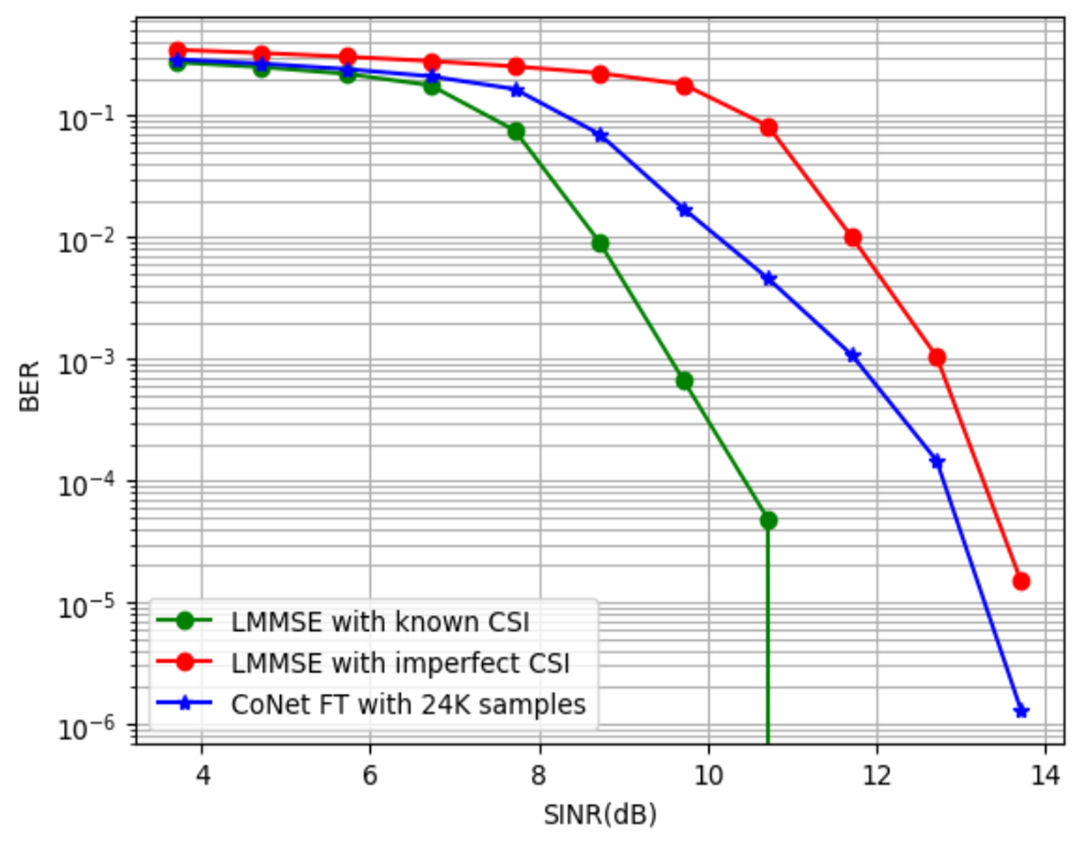}
        \label{fig:3}
    }
    \hfill
    \subfigure[The receiver is updated by 50K samples from the shifted distribution.]{
        \includegraphics[width=0.7\linewidth]{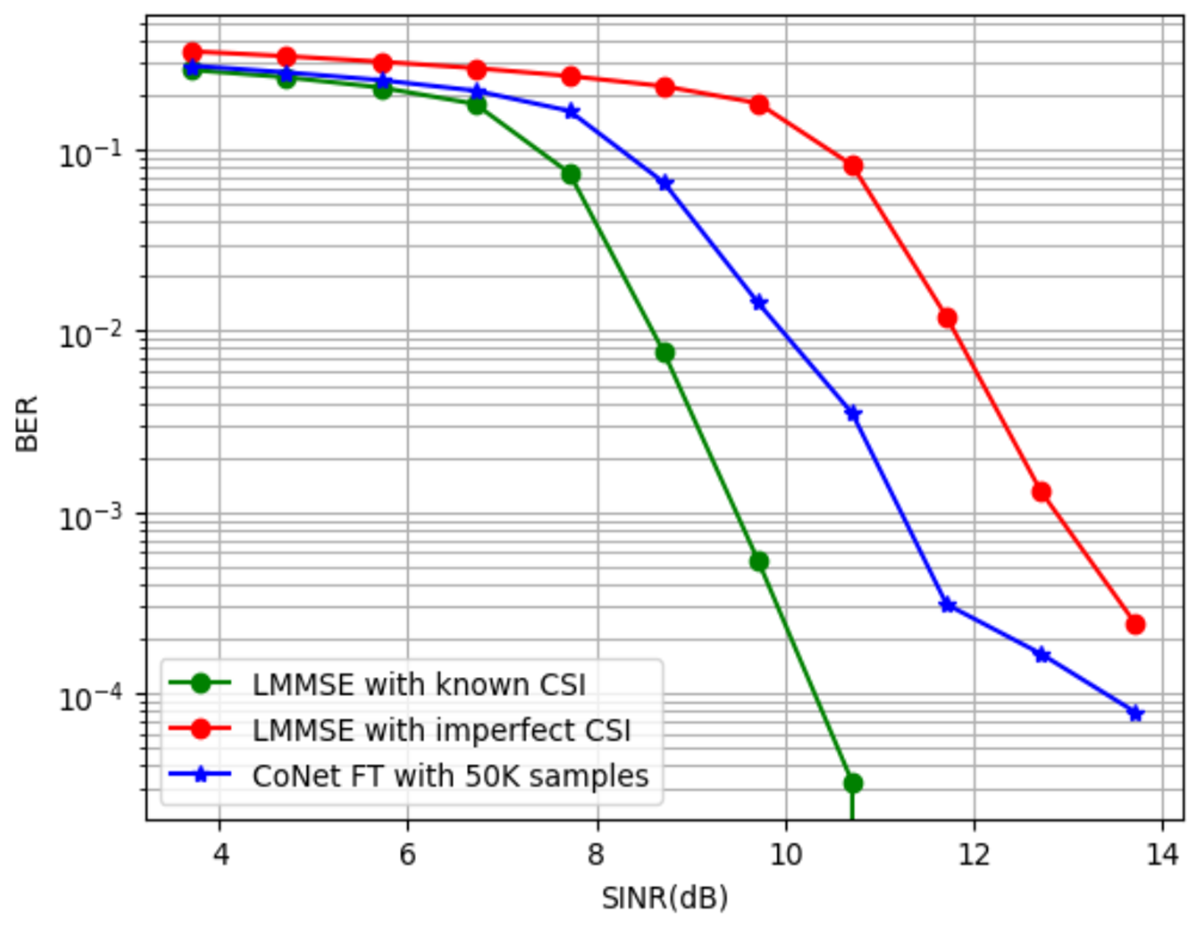}
        \label{fig:4}
    }
    \hfill
    \caption{BER versus SNR to show how the fine-tuned (FT) models track the channel distribution shift.}
    \label{fig:combined2}
\end{figure}
This section presents the simulation results of the proposed online
learning approach, examining its ability to track fast channel
distribution changes and assessing its impact on the BER.
Table~\ref{tab:settings} collects the complete system,
channel, receiver and training configuration used throughout, so that the
results below can be reproduced. The simulated uplink system consists of
a single-antenna user transmitting to a base station equipped with two
receive antennas. Channel realizations are generated using the
3GPP-defined tapped-delay-line (TDL) model, and distribution shifts are
simulated by varying the delay spread range and the user velocity. The pilot symbols and
their placements change from mini-batch to mini-batch. Sionna, a
Python-based link-level simulation library \cite{hoydis2022sionna}, is
used to model the system.

Two types of neural receivers are considered: the ResNet-based receiver
proposed in \cite{honkala2021deeprx} and the collaborative neural
networks (CoNet) architecture introduced in \cite{conet}. Although both
perform well with the proposed approach, CoNet shows faster adaptation
and better overall performance, and the results presented here therefore
focus on CoNet. The model is deliberately compact, with approximately
$600$k trainable parameters, reflecting the need for fast inference and
low memory usage in the physical layer. It is first trained offline on
20M samples, one epoch sufficing for convergence, and then fine-tuned
online with the pilot designs of Section~\ref{Sec:Pilots}. Training uses
the fully randomized design of Fig.~\ref{fig:OFDM1} throughout, as it
generalizes well to the other configurations in Fig.~\ref{fig:Pilots}.

Where a parameter was selected empirically rather than
fixed by the system model, Table~\ref{tab:settings} reports the range
examined alongside the chosen value; this applies to the masking ratio and to the learning rate, which
was swept independently for each method so that every scheme is evaluated
at its own best setting. The random seeds are fixed and shared across the
receivers being compared, so that at every mini-batch all receivers
observe the same channel realization, the same noise realization and the
same network input; any difference between the reported curves is
therefore attributable to the method rather than to realization noise.

We first compare the performance of the trained neural receiver against the traditional LMMSE receiver under both matched and shifted channel distributions. Fig.~\ref{fig:1} shows that when the model is tested on unseen samples drawn from the same distribution as the training set, the neural receiver significantly outperforms the LMMSE receiver with imperfect channel state information (CSI) and performs close to the infeasible LMMSE receiver with perfect CSI. However, when the channel distribution shifts (for example, by increasing the delay spread to the range [400, 410] ns), Fig.~\ref{fig:2} shows that the performance of the fixed neural receiver degrades sharply, particularly at high SNR values. This confirms the well-known vulnerability of neural receivers to distribution shifts.  

\begin{figure}
    \centering
    \includegraphics[width=0.7\linewidth]{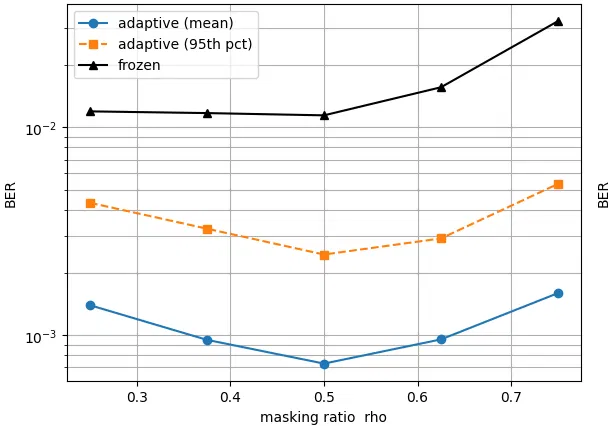}
    \caption{Sensitivity to the masking ratio $\rho$ under
the delay-spread variation of Fig.~\ref{fig:7}.}
    \label{fig:rho_sweep}
\end{figure}


To address this, we apply the proposed online fine-tuning method. With 24k adaptation samples drawn from the shifted distribution, the BER improves considerably as shown in Fig.~\ref{fig:3}. Moreover, with 50k samples, the receiver approaches to its original performance as shown in Fig.~\ref{fig:4}. These results demonstrate the effectiveness of online learning in recovering from sudden distribution shifts.  

We next examine how the receiver adapts continuously to gradual and fast channel variations, focusing on the delay analysis cases described in Section~\ref{Sec:Delay} and the analytical analysis.

Fig.~\ref{fig:rho_sweep} reports the sensitivity of the scheme to the
masking ratio $\rho$, swept over $\{0.25,\dots,0.75\}$, corresponding to
$32$ to $96$ masked pilots out of $128$. Each configuration is evaluated
twice, once with adaptation enabled and once with the model frozen.

The results show a clear trade-off with an interior optimum at
$\rho = 0.5$, and the two opposing mechanisms of
Section~\ref{sec:rho} are both visible. Increasing $\rho$ reduces the
number of pilots available for demodulation, and the frozen curve
degrades accordingly, rising by a factor of $2.35$ between $\rho = 0.5$
and $\rho = 0.75$. Decreasing $\rho$ reduces the number of labels
available per adaptation step, increasing the variance of the online
gradient; consequently the adaptation gain over the frozen receiver falls
from $15.7\times$ at $\rho = 0.5$ to $8.6\times$ at $\rho = 0.25$. The
minimum BER, $7.3 \times 10^{-4}$, occurs where these two effects
balance.

The optimum is broad. Every ratio in $0.375 \le \rho \le 0.625$ attains a
BER within $1.31\times$ of the minimum, so the scheme is not sensitive to
this choice and $\rho = 0.5$ is not a knife-edge setting. We adopt
$\rho = 0.5$ throughout as the empirically validated value, consistent
with the $\rho^\star = 0.47$ obtained by fitting \eqref{eq:excess} in
Section~\ref{sec:rho}.



\begin{figure}[t!]
    \centering
    
    \subfigure[The delay spread mean change versus the number of mini-batches.]{
        \includegraphics[width=0.75\linewidth]{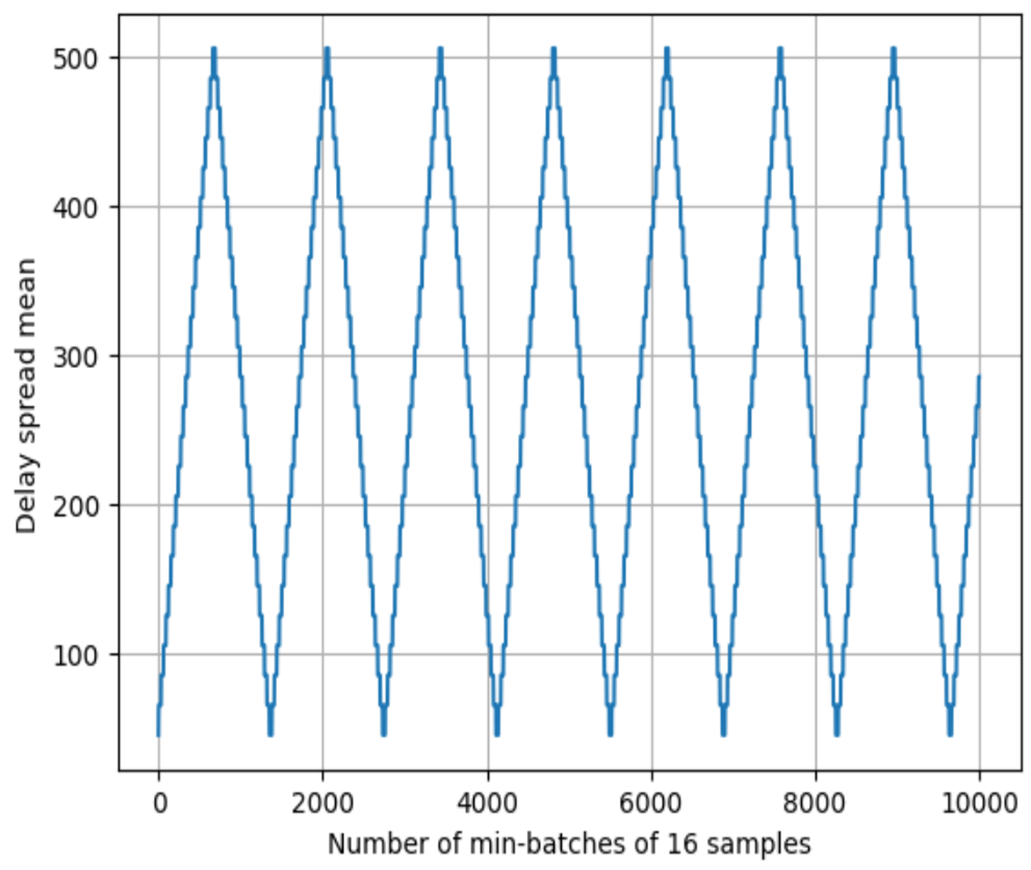}
        \label{fig:7}
    }
    \hfill
    \subfigure[The BER performance versus the number of received mini-batches in Case II ($U=2$).]{
        \includegraphics[width=0.8\linewidth]{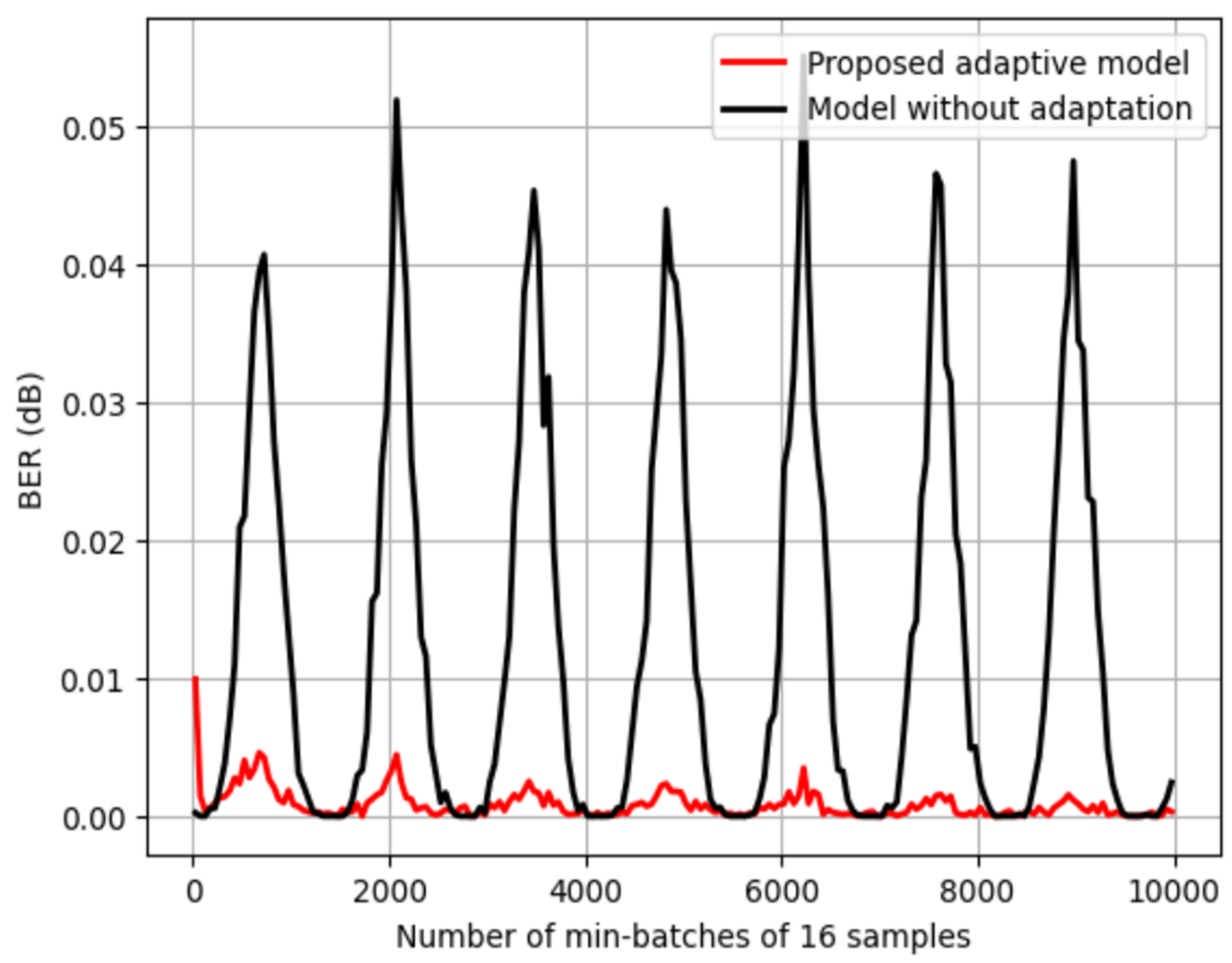}
        \label{fig:8}
    }
    \hfill
  
    \subfigure[The BER performance versus the number of received mini-batches in Case III ($U = 3$).]{
        \includegraphics[width=0.8\linewidth]{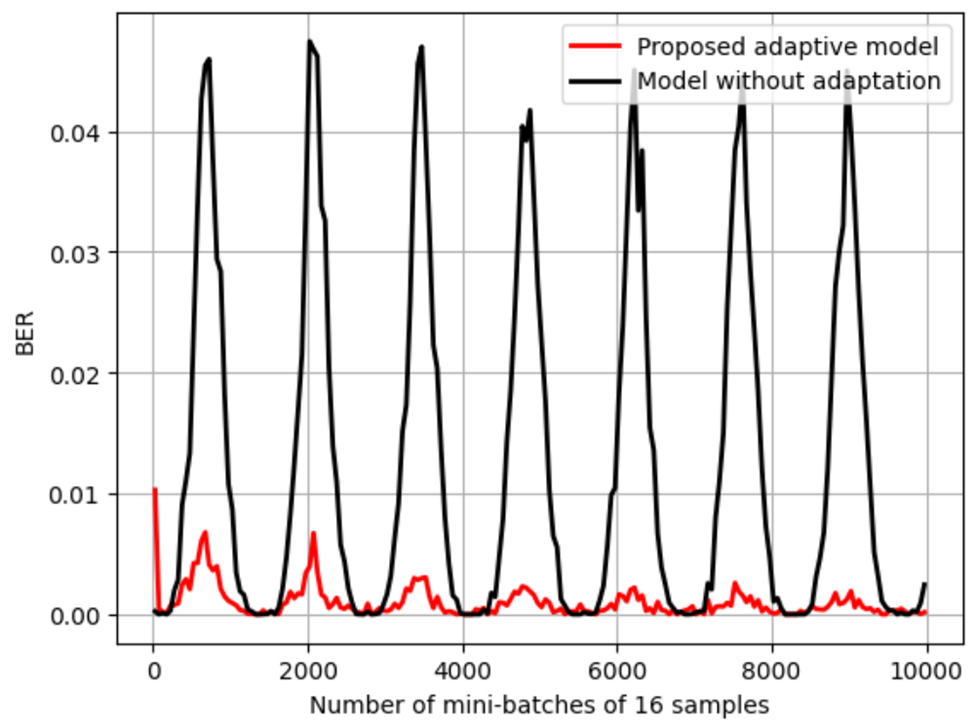}
        \label{fig:9}
    }
    \hfill
    
    \caption{Adaptation under a fast-rate
delay-spread shift. (a) The delay-spread profile, which increases by
$20$~ns every $30$ mini-batches ($480$ samples) and is common to both
cases. (b) Case~II: the receiver is updated every $U=2$ mini-batches.
(c) Case~III: the receiver is updated every $U=3$ mini-batches. The
channel evolution is identical in (b) and (c); only the update interval
differs, so the comparison isolates the effect of a longer update
interval.}
    \label{fig:combined3}
\end{figure}


We consider that the channel distribution shift is rapid, with the mean delay spread increasing by 20 ns every 480 samples, as illustrated in Fig.~\ref{fig:7}. Panels~(b) and~(c) of Fig.~\ref{fig:combined3} are driven by the identical
delay-spread profile of panel~(a) and differ only in the update interval,
$U=2$ and $U=3$ respectively. Their comparison therefore isolates the
cost of updating less frequently, which is the quantity analyzed in
Section~\ref{Sec:Delay}. Fig.~\ref{fig:8} shows the corresponding BER performance of both models under this faster variation with $U=2$. The results indicate that the proposed online approach continues to track the distribution changes effectively, preventing the receiver from experiencing severe degradation. Moreover, the performance of the adaptive model steadily improves as the number of adaptation cycles increases.

In Case~III, where the receiver updates less frequently, the proposed approach still adapts successfully and prevents catastrophic errors as shown in Fig.~\ref{fig:9}. Comparing Fig.~\ref{fig:8} to Fig.~\ref{fig:9}, the performance in Case III is slightly worse than Case~II during the early stages, but it converges with additional mini-batches, showing only a minor degradation compared to the more frequent update scheme.

\begin{figure}[t!]
    \centering
    \subfigure[Random change of the delay spread.]{
        \includegraphics[width=0.75\linewidth]{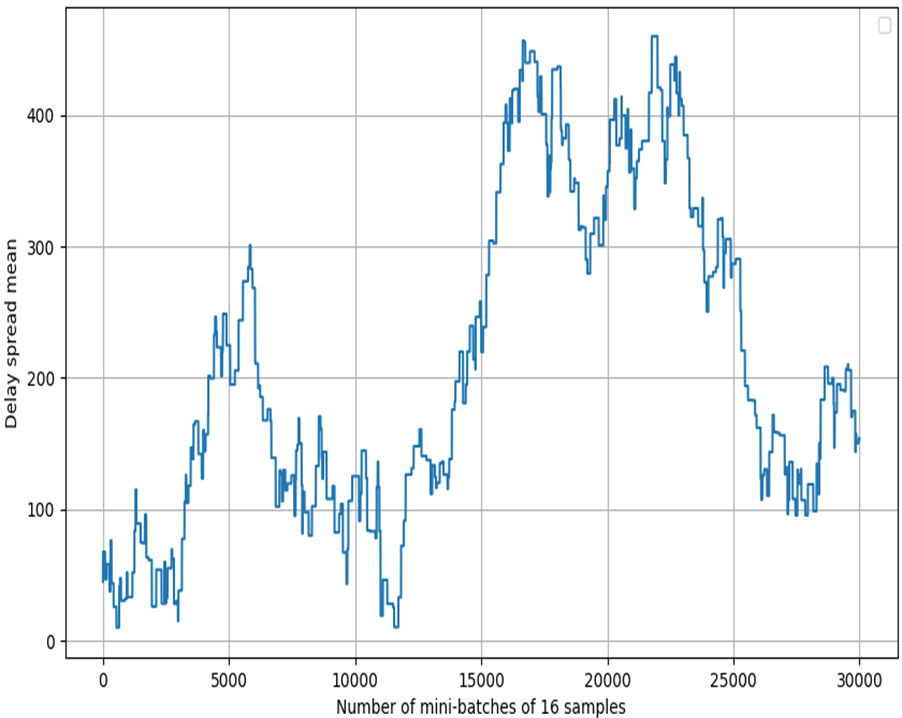}
        \label{fig:10}
    }
    \hfill
    \subfigure[The BER performance versus the number of received mini-batches.]{
        \includegraphics[width=0.8\linewidth]{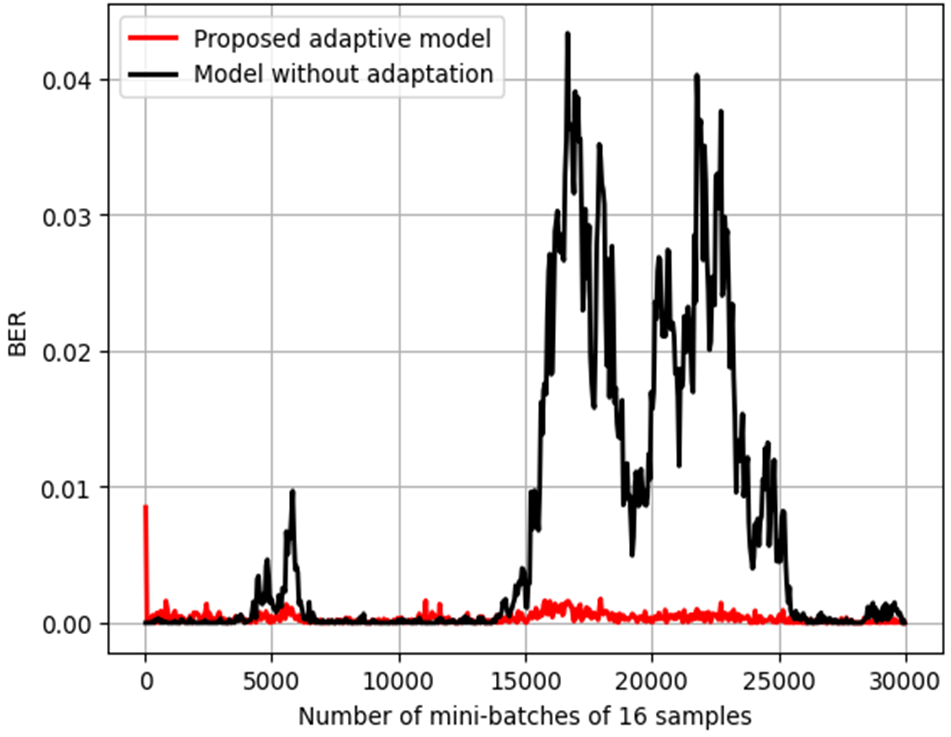}
        \label{fig:11}
    }
    \hfill
   \caption{The results of Case III, where the model is updated every third mini-batch with random-rate channel distribution shift.}
    \label{fig:combined4}
\end{figure}

Finally, consider the case where the mean delay spread varies randomly, following a distribution $\mathcal{U}(0,40)$ ns, with changes occurring at random intervals drawn from $\mathcal{U}(160,2400)$ samples. Under Case III, Fig.~\ref{fig:10} illustrates an example of how the delay spread evolves over time according to these distributions. The direction of each change, whether up or down, is random but exhibits some degree of correlation. As shown in Fig.~\ref{fig:11}, the proposed online learning approach maintains a consistently low BER under these conditions, whereas the fixed neural receiver suffers significant degradation.

Overall, the results show that the proposed online learning framework allows neural receivers to adapt to both slow and fast channel variations, preventing performance degradation by reusing existing pilots without dedicated training resources or service interruption. Channel distribution changes can be induced by variations in delay spread, user mobility, or considering different channel models. While the proposed online learning framework is capable of adapting to all these types of changes, we focus on delay spread variations in this work, as they have the most significant impact on BER performance.


The results of Figs.~\ref{fig:combined3} and \ref{fig:combined4} show that the
receiver tracks a continuously varying channel, but they do not quantify
how many samples are needed to restore performance after a change. That
quantity cannot be read from those figures, because the delay spread
there varies continuously and the receiver never reaches a steady state:
what they show is tracking rather than convergence. We therefore adopt a
separate protocol. The receiver begins from its pre-trained weights, the
delay spread is increased once by an amount $\Delta$ and then held
constant, and the BER is recorded until it settles. $E_b/N_0$ is fixed at $8$~dB so that the recovery curve is not blurred by SNR
variation, and each result is averaged over three independent runs. We
report the number of adaptation samples required for the BER to fall
below two fixed targets, $10^{-3}$ and $10^{-4}$, expressed in
mini-batches and in elapsed time at $0.5$~ms per slot.

We restrict the study to steps of at most $250$~ns. The delay spread is a
large-scale parameter that evolves as a user traverses a scattering
environment, and an abrupt change substantially larger than this is not
representative of practical propagation conditions. The continual-learning
results of Figs.~\ref{fig:combined3} and~\ref{fig:combined4} do span the full
$50$--$500$~ns range, but through gradual variation rather than a single
step.
\begin{table}[!t]
\renewcommand{\arraystretch}{1.2}
\caption{Recovery Sample Complexity}
\label{tab:samplecomplexity}
\centering
\scriptsize
\begin{tabular}{@{}lccccc@{}}
\toprule
& \multicolumn{3}{c}{\textbf{BER}}
& \multicolumn{2}{c}{\textbf{Mini-batches to reach}} \\
\cmidrule(lr){2-4}\cmidrule(lr){5-6}
$\Delta$ (ns) & no adapt. & at $t=0$ & settled & $10^{-3}$ & $10^{-4}$ \\
\midrule
$50$  & $<10^{-5}$ & $<10^{-5}$ & $2.3\times10^{-4}$
      & \multicolumn{2}{c}{not required} \\
$150$ & $1.9\times10^{-3}$ & $1.0\times10^{-3}$ & $1.6\times10^{-4}$
      & $25$ & $35$ \\
$250$ & $1.0\times10^{-2}$ & $3.6\times10^{-3}$ & $4.7\times10^{-4}$
      & $70$ & $275$ \\
\bottomrule
\end{tabular}

\vspace{2pt}
\begin{minipage}{\columnwidth}
\footnotesize
One mini-batch is $16$ frames, i.e.\ $8$~ms at $0.5$~ms
per slot. $E_b/N_0 = 8$~dB, averaged over three independent runs. BER
values below $10^{-5}$ are at the resolution limit of the measurement.
For $\Delta=50$~ns the pre-trained receiver is already error-free and no
recovery is required. The fixed receiver does not recover for
$\Delta \ge 150$~ns at any sample budget.
\end{minipage}
\end{table}

Table~\ref{tab:samplecomplexity} reports the results. For
$\Delta=150$~ns the BER returns below $10^{-3}$ within $25$ mini-batches,
corresponding to $400$ frames or $200$~ms, and below $10^{-4}$ within
$35$ mini-batches ($280$~ms). It settles at $1.6\times10^{-4}$, against
$1.9\times10^{-3}$ for a receiver that does not adapt, an improvement of
approximately $12\times$. For $\Delta=250$~ns the corresponding figures
are $70$ mini-batches ($560$~ms) and $275$ mini-batches ($2.2$~s), and
the BER settles at $4.7\times10^{-4}$ against $1.0\times10^{-2}$ without
adaptation, an improvement of approximately $21\times$. The sample cost
grows with the magnitude of the change, as expected.

Two observations follow. First, these recovery times are short relative
to the timescale on which the delay spread itself evolves, which is of
the order of seconds as a user moves through a scattering environment.
There is therefore a wide margin between the speed of adaptation and the
speed of the variation being tracked. Second, and more importantly, the
fixed receiver does not recover at any sample budget: its BER remains at
the post-change level indefinitely. The relevant comparison is thus not
only how quickly the proposed receiver recovers, but that it recovers at
all.

The case $\Delta=50$~ns is included for a different reason. Here the
pre-trained receiver is already error-free at this SNR and no recovery is
required, yet continuous adaptation introduces a small residual error of
order $10^{-4}$ where none was present. This is the gradient-noise floor
identified in Proposition~\ref{prop:tracking}: with a constant step size
the receiver does not converge to the optimum but fluctuates about it
with a steady-state error of order $\eta L \sigma_g^2$. When the channel
is not drifting, the tracking-lag term of \eqref{eq:tracking} vanishes
and this floor is all that remains, so adaptation incurs a small cost
without a corresponding benefit. This indicates that a deployed receiver
should gate its updates on a drift detector, applying them only when the
observed loss indicates that the channel statistics have changed, rather
than adapting unconditionally. The same effect can be reduced by lowering
the step size, at the expense of slower recovery when a change does
occur, a trade-off made explicit by the two competing terms in
\eqref{eq:tracking}.

\begin{figure}
    \centering
    \includegraphics[width=0.8\linewidth]{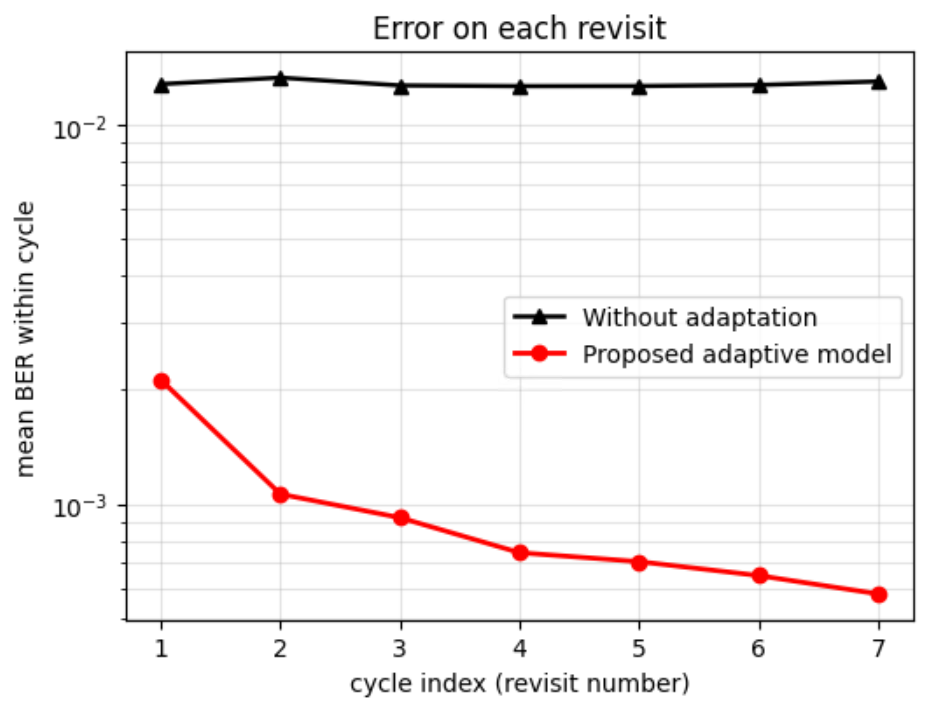}
    \caption{Mean BER within each traversal of the periodic
delay-spread schedule of Fig.~\ref{fig:7}, which revisits every
delay spread once per cycle.}
    \label{fig:forget}
\end{figure}

In our setting, the receiver performs the same task while only the channel distribution changes. Therefore, catastrophic forgetting would appear if the receiver performed poorly when a previously seen channel condition reappears. The periodic channel variation in Fig.~\ref{fig:7} allows us to evaluate this directly. As shown in Fig.~\ref{fig:forget}, the adaptive receiver improves across the seven cycles, with the error decreasing from $1.9\times10^{-3}$ to $6.2\times10^{-4}$. In contrast, the non-adaptive receiver remains near $1.3\times10^{-2}$. These results indicate that the adaptive receiver does not suffer from catastrophic forgetting and instead continues to improve as previously seen channel conditions are revisited.

Continual-learning methods such as elastic weight consolidation (EWC) and experience replay are designed to reduce forgetting, but they still require labeled data for online updates. Our method is complementary, as it provides these labels during deployment through the proposed pilot design. Since no significant forgetting is observed in Fig.~\ref{fig:forget}, we do not introduce the additional complexity of these methods. Their combination with the proposed framework can be investigated in future work if forgetting becomes significant.

\begin{figure}
    \centering
    \includegraphics[width=0.8\linewidth]{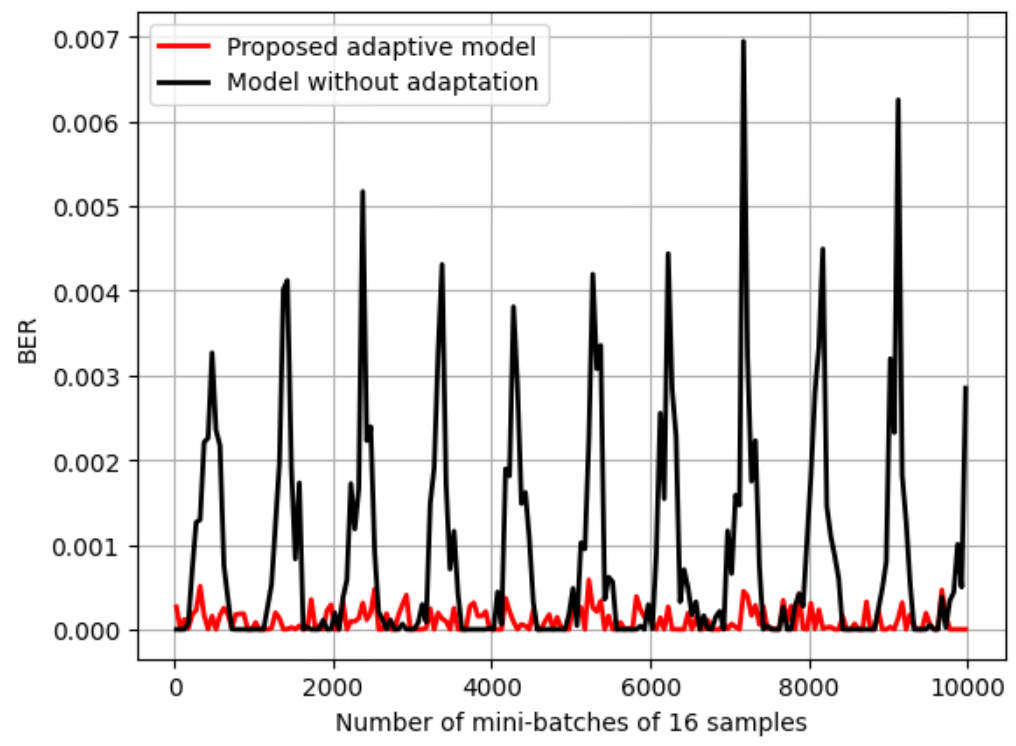}
    \caption{BER under time-varying user velocity. The
velocity follows the profile of Fig.~\ref{fig:7}, sweeping between
$0$ and $30$~m/s.  In
the low-velocity intervals the adaptive receiver shows a small residual
error absent from the fixed receiver, corresponding to the gradient-noise
floor of Proposition~\ref{prop:tracking}.}
\label{fig:doppler}
\end{figure}

\subsection{Time-Varying User Velocity}
\label{sec:doppler}

Delay-spread variation alters the frequency-domain correlation of the
channel. To confirm that the mechanism is not specific to that form of
non-stationarity, we repeat the experiment with a time-varying user
velocity, which alters the time-domain correlation instead. The velocity
follows the same profile as the delay spread in Fig.~\ref{fig:7},
sweeping periodically between $0$ and $30$~m/s in place of the
$50$--$500$~ns range, corresponding to a maximum Doppler
shift of approximately $350$~Hz at a carrier frequency of $3.5$~GHz, so
that the two experiments differ only in which channel parameter is
varied.

Fig.~\ref{fig:doppler} shows the resulting BER. The receiver without
adaptation degrades at every velocity peak, reaching
$7.0\times10^{-3}$, whereas the adaptive receiver remains close to
$2\times10^{-4}$ throughout, an improvement of between $15$ and $30$
times at the peaks. Since velocity and delay spread stress different axes
of the channel correlation, this indicates that the proposed mechanism
responds to distribution shift in general rather than to a particular
channel parameter.

The figure also illustrates a limitation of unconditional adaptation. In
the intervals between the velocity peaks, where the fixed receiver is
essentially error-free, the adaptive receiver exhibits a small residual
error of order $10^{-4}$. This is the gradient-noise floor of
Proposition~\ref{prop:tracking}: with a constant step size the receiver
fluctuates about the optimum rather than settling on it, and when the
channel is stationary the tracking-lag term of \eqref{eq:tracking}
vanishes and this floor is all that remains. The same behavior appears
for small delay-spread steps in Table~\ref{tab:samplecomplexity}. A
deployed receiver should therefore gate its updates on a drift detector,
adapting only when the observed loss indicates a change in the channel
statistics; reducing the step size has the same effect at the expense of
slower recovery, a trade-off made explicit by the two competing terms of
\eqref{eq:tracking}.

\subsection{Comparison with an Online Learning Baseline}
\label{sec:ecc}

The comparisons presented so far are against LMMSE detection and a fixed
neural receiver. Neither adapts online, so they establish the value of
adaptation but not the value of the particular way in which the proposed
scheme obtains its labels. We therefore compare against ECC-based online
label recovery \cite{schibisch2018online}, which addresses the same
problem and, like the proposed scheme, consumes no dedicated pilot
resources.

\subsubsection{The baseline}

A neural receiver cannot fine-tune during deployment because the
transmitted data bits are unknown. A body of work addresses this by
recovering the labels from the channel code
\cite{schibisch2018online, 9242305, 8999801}:
a received block is demodulated and decoded, and if decoding succeeds,
the decoded information bits are re-encoded to reconstruct the
transmitted codeword. The pair consisting of the received signal and the
reconstructed codeword then serves as a labeled training example. The
approach was introduced in \cite{schibisch2018online}, developed as a
self-supervised online training method in \cite{8999801},
and applied to multiuser MIMO detection in \cite{9242305}.
Its appeal is the same as that of the present work: it requires no
dedicated pilot resources and no change to the frame structure, and it
therefore tracks channel variations without additional spectral overhead.
We adopt it as our baseline for precisely this reason. Following
\cite{9242305}, we gate the updates on successful decoding,
so that blocks which fail the integrity check are discarded rather than
used as training data.

Three costs follow from this construction. First, reconstructing the
codeword requires a channel encoder to be invoked in the receive path.
An encoder is not otherwise part of the receive chain, so its
presence in the adaptation loop is an additional functional dependency.
Second, the decoder is placed inside the adaptation loop: every update
requires a full decoding operation in addition to the forward and
backward passes, whereas in the proposed scheme the decoder is needed
only to deliver the payload and plays no part in adaptation. Third, and most consequentially, 
the accuracy of the labels is a function of the receiver's own performance. The premise of
\cite{schibisch2018online, 9242305} is that the error
correcting code recovers the transmitted message even when some symbol
errors are present, so that the recovered labels are correct despite an
imperfect receiver. This holds while the receiver is only mildly
mismatched. When the mismatch is large enough that decoding fails, the
premise no longer applies: the recovered labels are wrong, the update
degrades the receiver further, and the degradation is self-reinforcing.
The regime in which the method is most needed is therefore the regime in
which its labels are least reliable, a limitation also noted in
\cite[Sec.~IV-C]{schibisch2018online}. The proposed design has no such
dependence, since its labels are pilot symbols known to the receiver in
advance.

A further practical difference is timing. Decoding and re-encoding take
time, so the labels available at a given instant correspond to a block
received earlier. In a time-varying channel the receiver is therefore
adapting towards conditions that have already changed. The proposed scheme has no such delay, since the masked
pilots are known at the instant of reception.

To isolate the effect of the label source, the baseline is implemented
within the same framework as the proposed scheme. At every mini-batch
both receivers observe the same channel realization, the same noise
realization and the same network input, and both are updated with Adam
under identical gradient clipping and update interval. The learning rate
of each was selected independently over
$\{10^{-5},\dots,10^{-3}\}$ and each is reported at its own best value.

\subsubsection{Behavior under time-varying delay spread}

\begin{figure}
    \centering
    \includegraphics[width=0.9\linewidth]{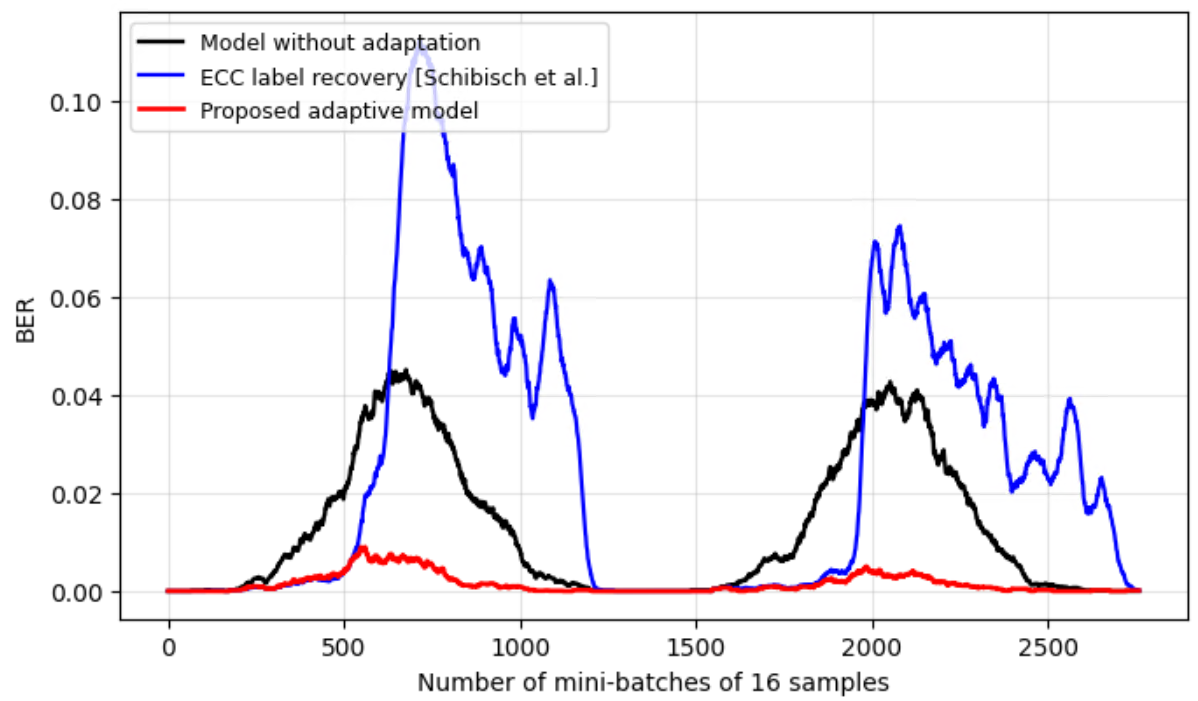}
    \caption{BER under the periodic delay-spread profile of
Fig.~\ref{fig:7}, with $E_b/N_0$ drawn uniformly on $[6,12]$~dB,
$U=2$ and code rate $0.5$. During the delay-spread excursions
the baseline's recovered labels are corrupted by decoding failures and
its BER exceeds that of the non-adaptive receiver.}
    \label{fig:ecc_timeseries}
\end{figure}

Fig.~\ref{fig:ecc_timeseries} shows the BER of the three receivers under
the periodic delay-spread profile of Fig.~\ref{fig:7}, with
$E_b/N_0$ drawn uniformly on $[6,12]$~dB and updates every $U=2$
mini-batches. During the intervals of small delay spread all three
receivers perform well and the two adaptive schemes are
indistinguishable. As the delay spread increases, the fixed receiver
degrades to a BER of $4.5\times10^{-2}$, as expected. The baseline,
however, degrades further still, reaching $1.1\times10^{-1}$: its
decoding failures corrupt the recovered labels, and the resulting
updates move the receiver away from rather than towards the correct
solution. It also recovers slowly once the delay spread falls again,
because the damage accumulated during the excursion must first be
undone. The proposed receiver remains below $9\times10^{-3}$ throughout.
The baseline is thus not merely less effective than the proposed scheme
in this regime; it is worse than performing no adaptation at all.

\begin{figure}
    \centering
    \includegraphics[width=0.9\linewidth]{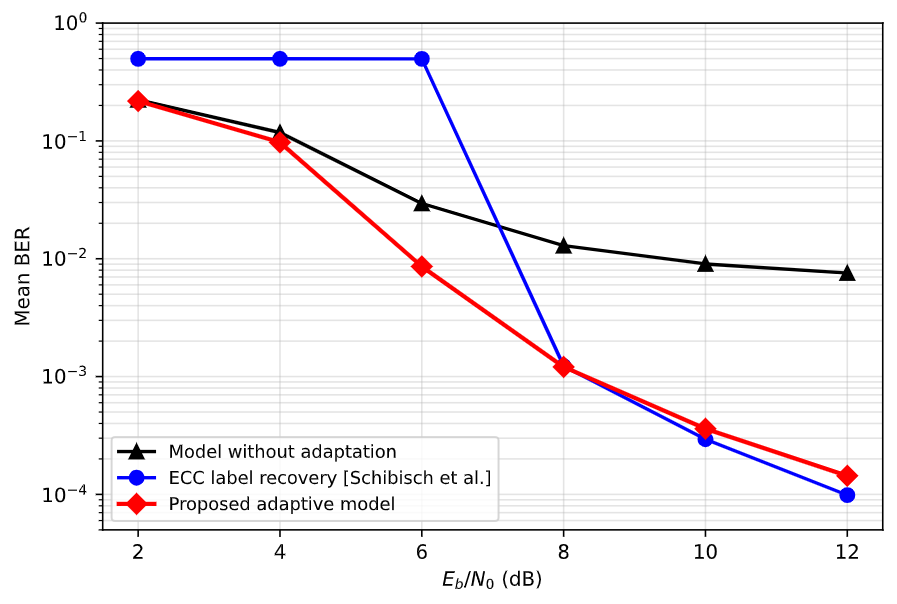}
    \caption{Mean BER against $E_b/N_0$ at a code rate of
$0.5$, under the time-varying delay-spread profile of
Fig.~\ref{fig:7}. The ECC baseline collapses below $8$~dB, where
decoding failures corrupt its recovered labels, and performs comparably
to the proposed scheme above $8$~dB, where decoding is reliable.}
    \label{fig:compare05}
\end{figure}
\subsubsection{Dependence on SNR and code rate}

Fig.~\ref{fig:compare05} resolves this behavior across SNR at a code
rate of $0.5$. Above $8$~dB the decoder succeeds on essentially every
block, the recovered labels are almost exact, and the baseline matches
the proposed scheme; at $10$ and $12$~dB it is in fact better by factors
of $1.2$ and $1.5$. We report this explicitly, as it is the regime in
which the baseline is strongest: it trains on every data-carrying
resource element and therefore receives roughly twelve times as many
labeled positions per sample as the proposed scheme, which uses only the
masked pilots. Below $8$~dB the picture reverses. Decoding failures
become frequent, the recovered labels degrade, and the baseline collapses
to a BER of approximately $0.5$, far worse than applying no adaptation at
all.

\begin{figure}
    \centering
    \includegraphics[width=0.9\linewidth]{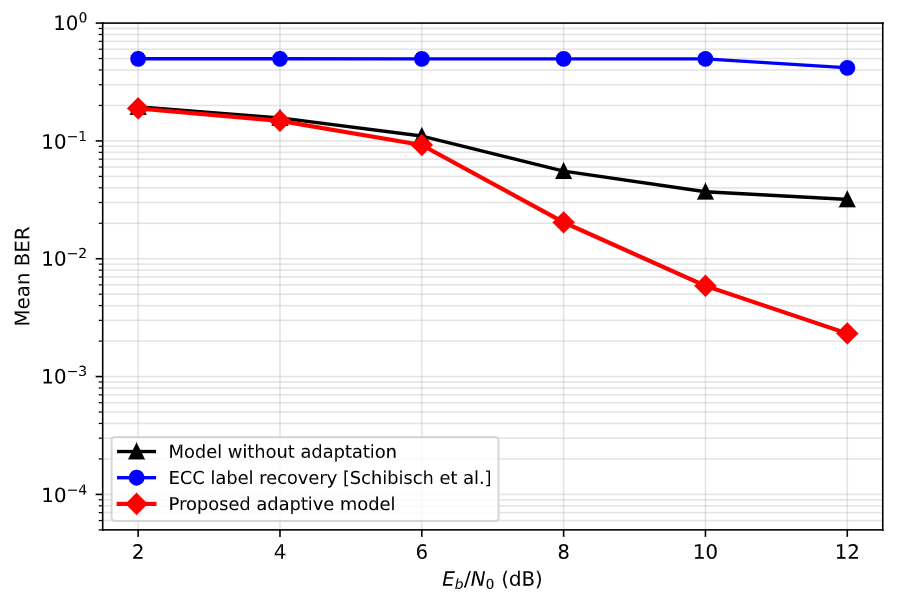}
    \caption{As Fig.~\ref{fig:compare05}, at a code rate of
$0.7$. The weaker code sustains decoding failures to much higher SNR, and
the ECC baseline does not recover anywhere in the range examined.}
    \label{fig:compare07}
\end{figure}

The code rate is a further determinant, and one the receiver does not
control, since the modulation and coding scheme is selected by link
adaptation and 5G~NR supports rates up to approximately $0.93$.
Fig.~\ref{fig:compare07} repeats the sweep at a rate of $0.7$. The
weaker code tolerates less mismatch, so decoding failures persist to much
higher SNR and the baseline does not recover anywhere in the range
examined, remaining above $0.4$ even at $12$~dB; the proposed scheme
outperforms it by between $2.6\times$ and $180\times$. The rate-$0.5$
results therefore represent the most favorable case for the baseline
rather than a typical one.

The two approaches are thus comparable when the decoder is reliable and
diverge sharply when it is not. The proposed design removes the
dependence of label quality on receiver state, on SNR and on code rate,
and does so without an encoder in the receive path or a decoder in the
adaptation loop. Its own gain over a non-adapting receiver grows
monotonically with SNR, reaching $52.6\times$ at $12$~dB at rate $0.5$
and $13.7\times$ at rate $0.7$.

\subsection{Uncertainty Quantification}
\label{sec:uq}

Table~\ref{tab:uq} quantifies the uncertainty of our
results at three levels, with $95\%$ intervals throughout. BER intervals
are computed over codewords, since LDPC decoding errors occur in bursts
within a codeword; a bit-level binomial interval would be up to $20$
times too narrow here.

The LLRs express the receiver's uncertainty about each
bit, and the decoder relies on them being calibrated. We measure
calibration by the expected calibration error (ECE) and by the
temperature $\tau$ by which the LLRs must be divided to minimize their
cross-entropy ($\tau>1$: over-confident). The pre-trained receiver is calibrated on the matched
channel. After the shift to $450$--$500$~ns its ECE grows $23$-fold and
$\tau=1.48$, so the decoder receives LLRs about $50\%$ larger than the
evidence supports; $500$ masked-pilot updates restore $\tau=0.99$ and raise
the bit-metric decoding (BMD) rate from $0.592$ to $0.685$. Adaptation
thus repairs the receiver's uncertainty estimates, not only its
decisions.

Since the network output does not depend on where the
masked pilots are placed, redrawing the masked-pilot set $500$ times on
each of $320$ received grids gives the exact sampling distribution of
$\widehat{\mathcal{L}}$. It is unbiased, and under i.i.d.\ placement its
variance matches $\sigma_\ell^2/(\rho P)$ of Lemma~\ref{lem:unbiased} to
within $1\%$ for $\rho P$ from $16$ to $256$. Our one-per-subcarrier
placement lowers the variance to $0.68$ of this value, so the lemma is
conservative; the gradient variance likewise decays as $1/(\rho P)$
(fitted slope $-1.01$). The receiver can therefore bound its own loss
from the masked pilots alone: the Student-$t$ interval built from the $\rho P$
per-element losses contained the full-grid loss in $96.8\%$ of cases at
a nominal $95\%$.

Finally, three independent repetitions of the Case~II
experiment of Fig.~\ref{fig:8} give a time-averaged BER of
$6.2\,[5.7,\,6.7]\times10^{-4}$ with adaptation and
$1.23\,[1.21,\,1.25]\times10^{-2}$ without, a ratio of
$19.9\,[18.5,\,21.3]$. A Bayesian treatment of the receiver weights,
e.g., with the Langevinized ensemble Kalman
filter~\cite{zhang2024lenkf,zhang2024elenkf,zhang2023covid}, would add
posterior uncertainty on the weights themselves and is left for future
work.

\begin{table}[!t]
\renewcommand{\arraystretch}{1.15}
\setlength{\tabcolsep}{3pt}
\caption{Uncertainty Quantification (95\% Confidence Intervals)}
\label{tab:uq}
\centering
\scriptsize
\begin{tabular}{@{}lccc@{}}
\toprule
& \textbf{40--50~ns} & \multicolumn{2}{c}{\textbf{450--500~ns}} \\
\cmidrule(lr){2-2}\cmidrule(lr){3-4}
& pre-trained & no adapt. & adapted \\
\midrule
\multicolumn{4}{@{}l}{\emph{Calibration of the LLRs}} \\
ECE & $0.0014$ & $0.0327$ & $0.0012$ \\
Temperature $\tau$ & $0.97$ & $1.48$ & $0.99$ \\
BMD rate (bit) & $0.770$ & $0.592$ & $0.685$ \\
& $[0.767,\,0.773]$ & $[0.585,\,0.599]$ & $[0.681,\,0.689]$ \\
\midrule
\multicolumn{4}{@{}l}{\emph{Coded BER at $E_b/N_0$ of}} \\
$7$~dB & $4.1\times10^{-5}$ & $6.0\times10^{-2}$ & $7.3\times10^{-3}$ \\
& $[0.7,\,13]\times10^{-5}$ & $[5.5,\,6.5]\times10^{-2}$ & $[5.6,\,9.3]\times10^{-3}$ \\
$10$~dB & $4.8\times10^{-6}$ & $3.1\times10^{-2}$ & $6.3\times10^{-4}$ \\
& $[0.1,\,27]\times10^{-6}$ & $[2.9,\,3.3]\times10^{-2}$ & $[4.1,\,9.3]\times10^{-4}$ \\
\midrule
\multicolumn{4}{@{}l}{\emph{Masked-pilot loss at $\rho P=64$ (Lemma~\ref{lem:unbiased})}} \\
Relative bias & \multicolumn{3}{c}{$+0.01\%$} \\
Var.\ ratio, i.i.d. & \multicolumn{3}{c}{$1.01\,[1.00,\,1.01]$} \\
Var.\ ratio, ours & \multicolumn{3}{c}{$0.68\,[0.66,\,0.70]$} \\
95\% CI coverage & \multicolumn{3}{c}{$96.8\%$} \\
\bottomrule
\end{tabular}

\vspace{2pt}
\begin{minipage}{\columnwidth}
\footnotesize
ECE: expected calibration error; $\tau>1$: over-confident LLRs; BMD rate:
$1-\mathbb{E}[\mathrm{BCE}]/\ln 2$ per coded bit; var.\ ratio: variance
of $\widehat{\mathcal{L}}$ over $\sigma_\ell^2/(\rho P)$; CI coverage:
fraction of grids whose full-grid loss lies in the pilot-only $95\%$
interval. Adapted: $500$ masked-pilot updates. $800$ codewords per
condition, $E_b/N_0\sim\mathcal{U}(7,10)$~dB; loss statistics from $500$
masked-pilot sets on each of $320$ grids at $450$--$500$~ns.
\end{minipage}
\end{table}

\subsection{Scaling to Wideband Carriers}
\label{sec:scaling}

The resource grid used above comprises $64$ subcarriers, whereas a 5G~NR
carrier may span several thousand. This reflects how we envisage the
receiver being deployed rather than a simplification. Since the receiver
is fully convolutional, its parameter count is independent of the FFT
size, so a wideband grid may be partitioned into $64$-subcarrier blocks,
each processed by the same network. Partitioning is valid because the
network exploits the \emph{local} correlation of the channel, set by the
coherence bandwidth rather than the carrier bandwidth: at $30$~kHz
spacing and a $500$~ns delay spread that bandwidth spans about ten
subcarriers, so a block already contains several coherence bandwidths and
a wider window would add little. It is also preferable to full-grid
processing. The computational cost is linear in the number of resource
elements either way, but partitioning bounds the activation memory of a
forward pass, and the blocks are independent and may be processed in
parallel, so inference latency is set by one block and does not grow with
carrier bandwidth\,---\,a binding constraint within a $0.5$~ms slot. The
results reported here thus correspond to one such block. Correlation
across block boundaries is not exploited, but since it extends only over
a coherence bandwidth the loss is confined to the edges and may be
reduced by overlapping adjacent blocks slightly. Scale also acts in
favor of the mechanism: a wideband carrier supplies proportionally more
pilots per slot, so by Lemma~\ref{lem:unbiased} the gradient variance is
lower and adaptation converges in fewer slots than reported here.

\section{Conclusion}
We presented an online and continual learning framework for neural OFDM
receivers that adapts to channel distribution shifts
without interrupting inference and, in two of its three pilot designs,
without consuming data-carrying resource elements. The key innovation lies in redesigning
demodulation pilots to serve a dual role, enabling both signal
demodulation and neural fine-tuning. We proposed three flexible pilot
structures and two receiver architectures that balance robustness and
computational complexity. We showed analytically that
the masked-pilot loss is an unbiased estimator of the full-grid loss,
which explains why ordinary fine-tuning remains stable here, and used its
variance to derive the optimal masking ratio and a tracking condition
relating the adaptation error to the channel drift rate and the update
interval. We also quantified the spectral, computational and
standardization overhead of the framework. Through delay analysis and
extensive simulations, we demonstrated that the proposed approach can
continuously track both slow and fast channel variations
in delay spread and user velocity, preventing the
catastrophic performance degradation observed in offline-trained neural
receivers. We further quantified the samples needed to
recover after a channel change, showed that the error decreases
monotonically across repeated traversals of the same conditions, and
found that online learning based on decoder-derived labels degrades below
a non-adapting receiver at low SNR or high code rate, where the proposed
scheme does not.
Future work includes extending the approach to multi-user MIMO systems,
validating it on real-world testbeds, and optimizing the online learning
approach hyperparameter (e.g., the mini-batch size, and the computation
resources).



\bibliographystyle{ieeetr}
\bibliography{Ref.bib}

@article{honkala2021deeprx,
  title={{DeepRx}: Fully convolutional deep learning receiver},
  author={Honkala, Mikko and Korpi, Dani and Huttunen, Janne MJ},
  journal={IEEE Transactions on Wireless Communications},
  volume={20},
  number={6},
  pages={3925--3940},
  year={2021},
  publisher={IEEE}
}

@inproceedings{korpi2021deeprx,
  title={{DeepRx MIMO}: Convolutional {MIMO} detection with learned multiplicative transformations},
  author={Korpi, Dani and Honkala, Mikko and Huttunen, Janne MJ and Starck, Vesa},
  booktitle={ICC 2021-IEEE International Conference on Communications},
  pages={1--7},
  year={2021},
  organization={IEEE}
}

@inproceedings{luostari2025adapting,
  title={Adapting to reality: Over-the-air validation of {AI-based} receivers trained with simulated channels},
  author={Luostari, Riku and Korpi, Dani and Honkala, Mikko and Huttunen, Janne MJ},
  booktitle={2025 IEEE Wireless Communications and Networking Conference (WCNC)},
  pages={01--06},
  year={2025},
  organization={IEEE}
}

@article{raviv2024adaptive,
  title={Adaptive and flexible model-based {AI} for deep receivers in dynamic channels},
  author={Raviv, Tomer and Park, Sangwoo and Simeone, Osvaldo and Eldar, Yonina C and Shlezinger, Nir},
  journal={IEEE Wireless Communications},
  volume={31},
  number={4},
  pages={163--169},
  year={2024},
  publisher={IEEE}
}

@article{khani2020adaptive,
  title={Adaptive neural signal detection for massive {MIMO}},
  author={Khani, Mehrdad and Alizadeh, Mohammad and Hoydis, Jakob and Fleming, Phil},
  journal={IEEE Transactions on Wireless Communications},
  volume={19},
  number={8},
  pages={5635--5648},
  year={2020},
  publisher={IEEE}
}

@article{jiang2021ai,
  title={{AI}-aided online adaptive {OFDM} receiver: Design and experimental results},
  author={Jiang, Peiwen and Wang, Tianqi and Han, Bin and Gao, Xuanxuan and Zhang, Jing and Wen, Chao-Kai and Jin, Shi and Li, Geoffrey Ye},
  journal={IEEE Transactions on Wireless Communications},
  volume={20},
  number={11},
  pages={7655--7668},
  year={2021},
  publisher={IEEE}
}

@inproceedings{schibisch2018online,
  title={Online label recovery for deep learning-based communication through error correcting codes},
  author={Schibisch, Stefan and Cammerer, Sebastian and D{\"o}rner, Sebastian and Hoydis, Jakob and ten Brink, Stephan},
  booktitle={2018 15th International Symposium on Wireless Communication Systems (ISWCS)},
  pages={1--5},
  year={2018},
  organization={IEEE}
}

@inproceedings{fischer2022adaptive,
  title={Adaptive neural network-based {OFDM} receivers},
  author={Fischer, Moritz Benedikt and D{\"o}rner, Sebastian and Cammerer, Sebastian and Shimizu, Takayuki and Lu, Hongsheng and Ten Brink, Stephan},
  booktitle={2022 IEEE 23rd International Workshop on Signal Processing Advances in Wireless Communication (SPAWC)},
  pages={1--5},
  year={2022},
  organization={IEEE}
}

@article{hoydis2022sionna,
  title={Sionna: An open-source library for next-generation physical layer research},
  author={Hoydis, Jakob and Cammerer, Sebastian and Aoudia, Fay{\c{c}}al Ait and Vem, Avinash and Binder, Nikolaus and Marcus, Guillermo and Keller, Alexander},
  journal={arXiv preprint arXiv:2203.11854},
  year={2022}
}

@inproceedings{conet,
  title={{CoNet-Rx}: Collaborative Neural Networks for {OFDM} Receivers},
  author={Mohanad Obeed and Ming Jian},
  booktitle={2025 Globecom},
  pages={1--6},
  year={2025},
  organization={IEEE}
}

@article{ye2017power,
  title={Power of deep learning for channel estimation and signal detection in {OFDM} systems},
  author={Ye, Hao and Li, Geoffrey Ye and Juang, Biing-Hwang},
  journal={IEEE Wireless Communications Letters},
  volume={7},
  number={1},
  pages={114--117},
  year={2017},
  publisher={IEEE}
}

@ARTICLE{9207745,
  author={Wei, Xiuhong and Hu, Chen and Dai, Linglong},
  journal={IEEE Transactions on Communications}, 
  title={Deep Learning for Beamspace Channel Estimation in Millimeter-Wave Massive MIMO Systems}, 
  year={2021},
  volume={69},
  number={1},
  pages={182-193},
  doi={10.1109/TCOMM.2020.3027027}}

@article{farahani2021brief,
  title={A brief review of domain adaptation},
  author={Farahani, Abolfazl and Voghoei, Sahar and Rasheed, Khaled and Arabnia, Hamid R},
  journal={Advances in data science and information engineering: proceedings from ICDATA 2020 and IKE 2020},
  pages={877--894},
  year={2021},
  publisher={Springer}
}

@article{obeed2025hybrid,
  title={Hybrid Neural/Traditional {OFDM} Receiver with Learnable Decider},
  author={Obeed, Mohanad and Jian, Ming},
  journal={arXiv preprint arXiv:2509.18574},
  year={2025}
}

@article{luan2023channelformer,
  title={Channelformer: Attention based neural solution for wireless channel estimation and effective online training},
  author={Luan, Dianxin and Thompson, John S},
  journal={IEEE Transactions on Wireless Communications},
  volume={22},
  number={10},
  pages={6562--6577},
  year={2023},
  publisher={IEEE}
}

@article{xu2024learning,
  title={Learning to estimate: A real-time online learning framework for {MIMO-OFDM} channel estimation},
  author={Xu, Jiarui and Li, Lianjun and Zheng, Lizhong and Liu, Lingjia},
  journal={IEEE Transactions on Wireless Communications},
  year={2024},
  publisher={IEEE}
}

@article{du2022robust,
  title={Robust online {CSI} estimation in a complex environment},
  author={Du, Heng and Deng, Yun and Xue, Jiang and Meng, Deyu and Zhao, Qian and Xu, Zongben},
  journal={IEEE Transactions on Wireless Communications},
  volume={21},
  number={10},
  pages={8322--8336},
  year={2022},
  publisher={IEEE}
}

@article{zhang2024continuous,
  title={Continuous online learning-based {CSI} feedback in massive MIMO systems},
  author={Zhang, Xudong and Wang, Jintao and Lu, Zhilin and Zhang, Hengyu},
  journal={IEEE Communications Letters},
  volume={28},
  number={3},
  pages={557--561},
  year={2024},
  publisher={IEEE}
}

@article{cui2022unsupervised,
  title={Unsupervised online learning in deep learning-based massive {MIMO CSI} feedback},
  author={Cui, Yiming and Guo, Jiajia and Wen, Chao-Kai and Jin, Shi and Han, Shuangfeng},
  journal={IEEE Communications Letters},
  volume={26},
  number={9},
  pages={2086--2090},
  year={2022},
  publisher={IEEE}
}

@article{wang2022learn,
  title={Learn to adapt to new environments from past experience and few pilot blocks},
  author={Wang, Ouya and Gao, Jiabao and Li, Geoffrey Ye},
  journal={IEEE Transactions on Cognitive Communications and Networking},
  volume={9},
  number={2},
  pages={373--385},
  year={2022},
  publisher={IEEE}
}

@article{ghadimi2013,
author = {Ghadimi, Saeed and Lan, Guanghui},
title = {Stochastic First- and Zeroth-Order Methods for Nonconvex Stochastic Programming},
journal = {SIAM Journal on Optimization},
volume = {23},
number = {4},
pages = {2341-2368},
year = {2013},
doi = {10.1137/120880811},
URL = {       https://doi.org/10.1137/120880811},
eprint = {     https://doi.org/10.1137/120880811    }}

@article{simonetto2020,
  title={Time-varying convex optimization: Time-structured algorithms and applications},
  author={Simonetto, Andrea and Dall'Anese, Emiliano and Paternain, Santiago and Leus, Geert and Giannakis, Georgios B},
  journal={Proceedings of the IEEE},
  volume={108},
  number={11},
  pages={2032--2048},
  year={2020},
  publisher={IEEE}
}

@ARTICLE{9242305,
  author={Shlezinger, Nir and Fu, Rong and Eldar, Yonina C.},
  journal={IEEE Transactions on Wireless Communications}, 
  title={DeepSIC: Deep Soft Interference Cancellation for Multiuser MIMO Detection}, 
  year={2021},
  volume={20},
  number={2},
  pages={1349-1362},
  doi={10.1109/TWC.2020.3032663}}

@ARTICLE{8999801,
  author={Shlezinger, Nir and Farsad, Nariman and Eldar, Yonina C. and Goldsmith, Andrea J.},
  journal={IEEE Transactions on Wireless Communications}, 
  title={ViterbiNet: A Deep Learning Based Viterbi Algorithm for Symbol Detection}, 
  year={2020},
  volume={19},
  number={5},
  pages={3319-3331},
  doi={10.1109/TWC.2020.2972352}}

@misc{wang2021tent,
      title={Tent: Fully Test-time Adaptation by Entropy Minimization}, 
      author={Dequan Wang and Evan Shelhamer and Shaoteng Liu and Bruno Olshausen and Trevor Darrell},
      year={2021},
      eprint={2006.10726},
      archivePrefix={arXiv},
      primaryClass={cs.LG},
      url={https://arxiv.org/abs/2006.10726}, 
}

@book{goodfellow2016deep,
  author    = {Goodfellow, Ian and Bengio, Yoshua and Courville, Aaron},
  title     = {Deep Learning},
  publisher = {MIT Press},
  year      = {2016},
  note      = {Ch.~6.5}
}

@article{zhang2024lenkf,
  author  = {Zhang, Peiyi and Song, Qifan and Liang, Faming},
  title   = {A {L}angevinized Ensemble {K}alman Filter for Large-Scale Dynamic Learning},
  journal = {Statistica Sinica},
  volume  = {34},
  pages   = {1071--1091},
  year    = {2024}
}

@article{zhang2024elenkf,
  author  = {Zhang, Peiyi and Dong, Tianning and Liang, Faming},
  title   = {An Extended {L}angevinized Ensemble {K}alman Filter for Non-{G}aussian Dynamic Systems},
  journal = {Computational Statistics},
  volume  = {39},
  number  = {6},
  pages   = {3347--3372},
  year    = {2024},
  doi     = {10.1007/s00180-023-01443-4}
}

@article{zhang2023covid,
  author  = {Zhang, Peiyi and Dong, Tianning and Li, Ninghui and Liang, Faming},
  title   = {Identification of Factors Impacting on the Transmission and Mortality of {COVID}-19},
  journal = {Journal of Applied Statistics},
  volume  = {50},
  number  = {11--12},
  pages   = {2624--2647},
  year    = {2023},
  doi     = {10.1080/02664763.2021.1953449}
}

@inproceedings{obeed2026learning,
  title={Learning During Inference: Adapting Neural Wireless Receivers Via Demodulation Pilots},
  author={Obeed, Mohanad and Jian, Ming},
  booktitle={ICC 2026-IEEE International Conference on Communications},
  pages={1--6},
  year={2026},
  organization={IEEE}
}







\end{document}